\documentclass[
showpacs, floatfix, aps,prb,amsmath,
twocolumn, superscriptaddress,
eqsecnum ]{revtex4}

\usepackage{bm}
\usepackage{array}
\newcommand{\ocite}[1]{
[\onlinecite{#1}]}%\newcommand{\req}[1]{Eq.~(\ref{#1})}

\newcommand{\be}{\begin{equation}}
\newcommand{\ee}{\end{equation}}

\newcommand{\bea}{\begin{eqnarray}}
\newcommand{\eea}{\end{eqnarray}}

\newcommand{\bes}{\begin{split}}
\newcommand{\eess}{\end{split}}

\usepackage{amssymb}
\usepackage{latexsym}
\usepackage[dvips]{graphicx}
\usepackage{amsmath}
\usepackage[mathcal]{eucal}

\newcommand{\req}[1]{Eq.~(\ref{#1})}
\newcommand{\reqs}[1]{Eqs.~(\ref{#1})}
\newcommand{\rref}[1]{(\ref{#1})}
\newcommand{\oncite}[1]{Ref.~\onlinecite{#1}}
\renewcommand{\vec}[1]{{\bm #1}}

\newcommand{\vare}{\varepsilon}

\newcommand{\fnd}{F_{\rm d}}
\newcommand{\fni}{F_{\rm i}}

\newcommand{\wc}{\omega_{\rm c}}
\newcommand{\Tc}{T_{\mathrm{c}}}
\newcommand{\Rc}{R_{\mbox{\scriptsize c}}}
\newcommand{\vF}{v_{\rm F}}
\newcommand{\pF}{p_{\mathrm{F}}}
\newcommand{\Pmw}{{\cal P}_{\omega}}

\newcommand{\tautr}{\tau_{\rm tr}}
\newcommand{\tauin}{\tau_{\rm ee}}

\newcommand{\tauq}{\tau_{\rm q}}
\newcommand{\taul}{\tau_{\rm sm}}
\newcommand{\taus}{\tau_{\rm sh}}

\def\eac{\epsilon_{\mbox{\scriptsize ac}}}
\def\edc{\epsilon_{\mbox{\scriptsize dc}}}

\begin{document}
\title{Effect of microwave radiation on non-linear resistivity of a
two-dimensional electron gas at large filling factors}

\author{Maxim Khodas}

\affiliation{School of Physics and Astronomy, University of Minnesota, Minneapolis, MN 55455, USA }
\affiliation{Department of Condensed Matter Physics and Materials
Science, Brookhaven National Laboratory, Upton, NY 11973-5000, USA}

\author{Maxim G. Vavilov}

\affiliation{Department of Physics, University of Wisconsin, Madison, WI 53706, USA }

\date{October 30, 2008}
\begin{abstract}
We study transport properties of a two-dimensional electron gas,
placed in a classically strong perpendicular magnetic field and in
constant and oscillating in-plane electric fields. The analysis is
based on a quantum Boltzmann equation derived for a weakly
disordered two-dimensional electron gas.
We consider disordered potential with both long and short range
correlations. Electron scattering off such disorder is not limited
to small change in momentum direction, but occurs on an arbitrary
angle, including the backscattering.
The non-linearity of the transport in the considered system is a
consequence of two co-existing effects: formation of a
non-equilibrium distribution function of electrons and modification
of the scattering rate off the disorder in the presence of dc and ac
electric fields.
This work describes both effects in a unified way. The calculated
dissipative component of electric current oscillates as a function
of the electric field strength and frequency of microwave radiation
in a qualitative agreement with experiments.

\end{abstract}
\pacs{73.23.-b, 73.40.-c, 73.50.Fq}
\maketitle
\section{Introduction}
\label{sec:Intro}
The discovery of the microwave-induced resistance oscillations
(MIRO) \cite{ZDRS01,ye02} and the zero resistance states (ZRS)
\cite{mani02,ZDRS03} has raised interest in non-linear transport
properties of two-dimensional systems (2DES) in a perpendicular
magnetic field at large filling factors. The dissipative dc
magneto-resistance exhibits giant oscillations with the inverse
magnetic field when exposed to a microwave
radiation.\cite{ZDRS01,ye02,zudovPRL91,dorozhkin-2003,zudov04,smet05,studenikin05,du06}
The period of MIRO is controlled by the ratio of the microwave
frequency $\omega$ to the electron cyclotron frequency $\wc= \left|
e \right| B/m c$ in magnetic field $B$. In the high mobility samples
the MIRO evolve into the ZRS when the linear dc resistance becomes
negative.\cite{AAM03}
These remarkable findings made it imperative to understand the
mechanism of oscillations preceding the onset of the ZRS. The
experiments were performed at relatively high temperatures and low
magnetic fields, when the Shubnikov-de Haas oscillations are
suppressed.

The appearance of the MIRO has been first attributed to the
modification of the impurity scattering rates in the presence of the
magnetic field.\cite{ryzhii,durst03,VA03} This scenario is commonly
referred to as the ``displacement'' mechanism. It has been
predicted~\cite{dmitriev03,DVAMP} that a different, so-called
``inelastic'' mechanism, dominates in the regime, where both MIRO
and ZRS were observed.\cite{mani02,ZDRS03} According to
\oncite{DVAMP}, the microwave radiation is responsible for formation
of a non-equilibrium component of the distribution function, which
is isotropic in momentum and oscillates as a function of energy. The
amplitude of such non-equilibrium component of the distribution
function is characterized by the temperature dependent rate of
inelastic scattering processes $1/\tau_{\rm ee}$ due to the
electron-electron interaction. The analysis of \oncite{DVAMP}
suggests that in weak electric fields and at sufficiently low
temperatures the ``inelastic'' contribution from the non-equilibrium
component of the distribution function to the linear-response dc
resistivity is larger than the ``displacement'' contribution.

A different series of experiments focused on measurements of the
non-linear differential resistance in the absence of microwave
excitation.\cite{du02,vitkalov05,vitkalov06,zudovDC06,vitkalov07,zudov07}
In these experiments, the differential resistance has been measured
in the Hall bar geometry as a function of the applied direct
current. This current creates a strong electric field in a
perpendicular direction, known as the Hall field, provided that the
magnetic field is strong, $\wc\tau_{\rm tr}\gg 1$, where $\tautr$ is
the transport scattering time. The scattering off disorder in the
Hall field is accompanied by a change of electron kinetic energy and
leads to dependence of transport characteristics on the strength of
this field. In particular, the differential resistance exhibits
oscillations, called the Hall induced resistance oscillations
(HIRO), as a function of the Hall electric field strength $E$. The
HIRO were explained~\cite{du02} as a result of the geometric
resonance in the electron transitions between the tilted Landau
levels when the diameter of the cyclotron trajectory becomes
commensurable with the spatial modulation of the density of states.
More rigorous approach\cite{VAG} employing the quantum kinetic equation
showed that the ``inelastic''
mechanism is important in a relatively narrow interval of applied
electric fields and the ``displacement'' mechanism becomes dominant in
the regime of strong direct current, where HIRO were observed.

The effect of the microwave irradiation on the non-linear transport
was experimentally investigated in
Refs.~\onlinecite{zudovACDC07,zudovACDC08},
in which a
2DES was subject to both constant  and oscillating electric fields.
The value of the differential magneto-resistance depends on two
dimensionless parameters
\begin{equation}
\edc=\frac{|e| E (2\Rc)}{\wc},\quad \eac=\frac{\omega}{\wc}.
\label{eac_edc}
\end{equation}
where $E$ is the magnitude of the in-plane constant electric field,
$\Rc=v_{\rm F}/\wc$ is the cyclotron radius, $v_{\rm F}$ is the
Fermi velocity and $\wc$ is the cyclotron frequency;
throughout this paper we use $\hbar=1$.
Maxima of the magneto-resistance in the vicinity of the main
diagonal of the two dimensional $(\eac,\edc)$-plane are obtained
when the sum $\eac+\edc$ is integer.
Interestingly, this simple rule does not hold farther away from the
main diagonal $\edc\sim \eac$.
And in fact, the interplay between both types of excitation gives
rise to an unexpectedly rich structure of extremes and saddle points
of the differential magneto-resistance in the
$(\eac,\edc)$-plane.\cite{zudovACDC07}

Oscillations of the differential resistance as a function of $\edc$
are understood\cite{du02,vitkalov06,zudovDC06,VAG} in terms of
electron backscattering off impurities, which corresponds to change
of electron direction on its opposite. Therefore,  an appropriate
model of disordered potential has to include processes of electron
scattering on an arbitrary angle $\theta$, including $\theta=\pi$.
Such potential has both long range correlations being responsible
for small angle scattering, and short range correlations leading to
backscattering. The proper treatment of the disorder potential with
above properties requires a further extension of a kinetic theory of
2DES,\cite{VA03} developed for smooth disorder.

The goal of the present paper is to construct a systematic theory of
magneto-oscillations of the differential resistivity of the 2DES in
the presence of electric fields of arbitrary strength. In
experiments of Refs.~\onlinecite{zudovACDC07,zudovACDC08},
oscillations have been observed at large filling factors, $\sim
E_{\rm F}/ \wc\gg 1$ with $E_{\rm F}$ being the Fermi energy. In
this limit, we can treat the kinetics of electron gas
semiclassically. The analysis is performed in the experimentally
relevant range of classically strong magnetic fields, $ 1/ \tautr
\ll \wc $. In the present work we focus on the situation when Landau
levels are not resolved, which implies the inequality $\wc
\tauq\lesssim 1$, where $\tauq$ is the quantum scattering time.
The temperature is assumed to be relatively large, $ T \gtrsim
\wc $,  so that the Shubnikov-de Haas
oscillations are exponentially suppressed; $k_{\rm B}=1$.

The two most important contributions to the non-linear electric
current are the ``inelastic'' contribution originating from the
modification of the electron distribution
function~\cite{DVAMP,VAG,dorozhkin-2003,dmitriev07} and the
``displacement'' contribution representing the changes of electron
scattering amplitudes off disorder. Additional contributions were
identified and studied in \oncite{dmitriev07}, but these
contributions have additional smallness in systems with mixed
disorder. The displacement mechanism can be studied by various
methods,~\cite{ryzhii,durst03,VA03,xs03,leiacdc,KT07} which provide
a qualitatively correct picture for electron transport in strong
electric fields. However,  for a full description of the crossover
from "weak-" to "strong-" fields, the kinetic equation is necessary.

The paper is organized as follows. In Sec.~\ref{sec:Qualitative} we
present a simplified analysis of magneto-oscillations in combined
constant and oscillating electric fields and summarize the main
results. In Sec.~\ref{sec:Kin-Eq} the kinetic equation is derived in
the framework of the Keldysh formalism. We solve the kinetic
equation within a bilinear response in microwave field and apply
this solution to calculation of the non-linear current in
Sec.~\ref{sec:ACDC}. Sec.~\ref{sec:Arbitr-Power} contains an
analysis of the current beyond the bilinear in microwave field
response. Discussion and conclusions are presented in
Sec.~\ref{sec:Discussion}.
\section{Main results}
\label{sec:Qualitative}
\subsection{Bilinear response to the applied microwave radiation in
strong dc electric field}
\label{sec:Qualit-StrongDC}

\begin{figure}[h]
\includegraphics[width=0.7\columnwidth]{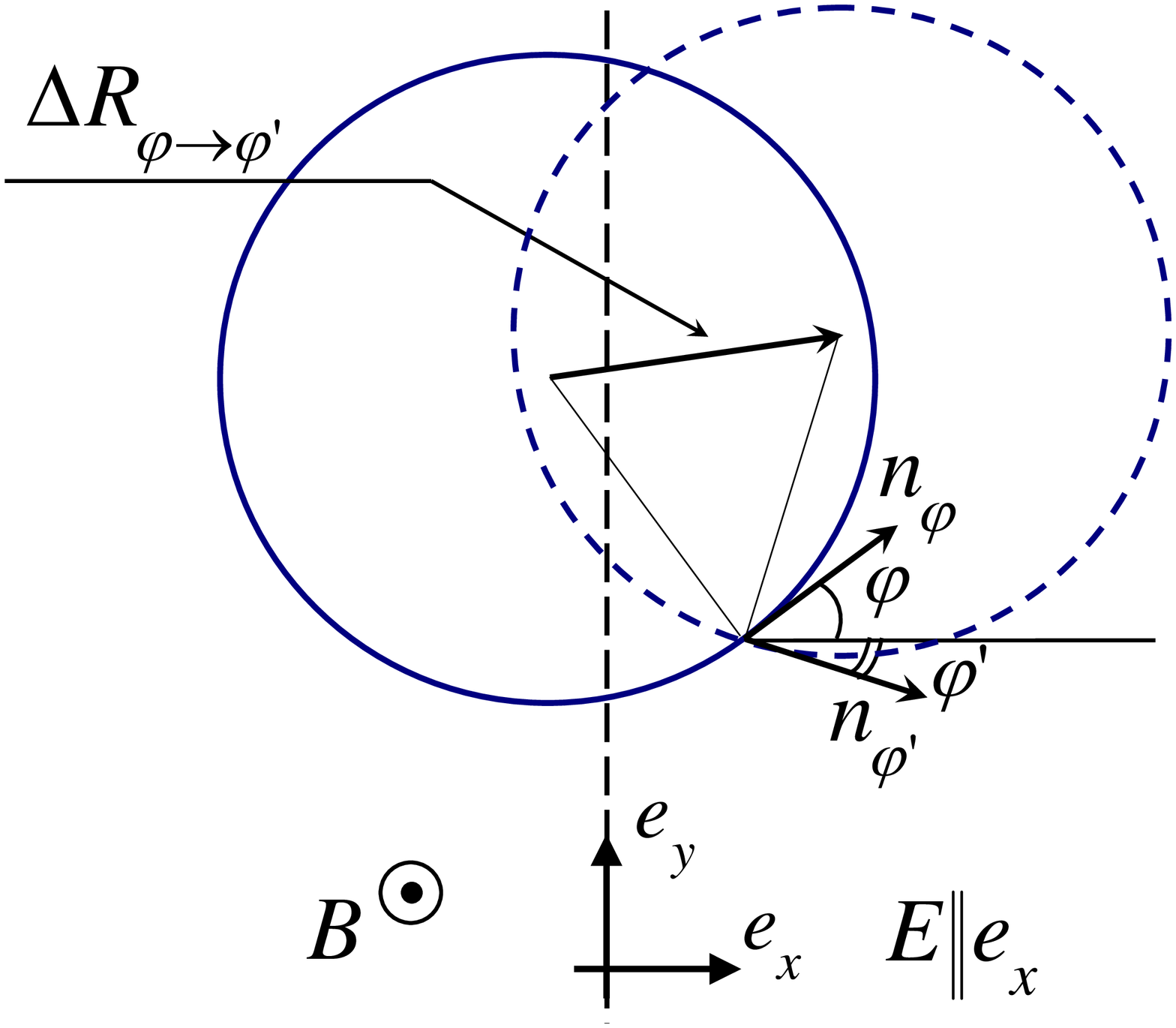}
%\vspace{-0.5cm}
\caption{Scattering off impurity leads to the shift of the guiding
center of a cyclotron electron trajectory. \label{fig:shift}}
\end{figure}

In this section we present heuristic discussion of the results of
the paper for the dissipative current in strong dc electric field,
but consider the contribution to the electric current that is only
bilinear in the microwave electric field.
Our analysis employs the semiclassical treatment of electron motion
in crossed electric and magnetic fields, valid in the limit of high
Landau level index $E_{\rm F}/\wc\gg 1$, where $E_{\rm F}$ is the Fermi
energy and $\wc$ is a cyclotron period. According to this picture
electron scattering off impurities amounts to the spatial shift of
the guiding center of the cyclotron orbit, Fig.~\ref{fig:shift}.
In the presence of the electric, $\vec{E} = \vec{e}_x E$, and magnetic, $\vec{B} = \vec{e}_z B$,
fields,  the
dissipative current results from the imbalance between the drift of
cyclotron orbits parallel to the electric field.
We denote a unit vector forming angle $\varphi$ with the
direction of the electric field $\vec{e}_x$ by
$\vec{n}_{\varphi}=\{\cos\varphi;\sin\varphi;0\}$.
The electron scattering resulting in the change of the direction of
motion from $\vec{n}_{\varphi}$ to $\vec{n}_{\varphi'}$ leads to the
shift of the guiding center $\Delta \vec{R}$ given by
\begin{equation} \label{qualit13}
\Delta \vec{R}_{\varphi \rightarrow \varphi'} = R_c \vec{e}_z \times
\left(\vec{n}_{\varphi} - \vec{n}_{\varphi'} \right) \, .
\end{equation}

We present the current as a sum of two contributions
\begin{equation} \label{qualit15}
j = j_{1} + j_{2}\, .
\end{equation}
Here, the first term
\begin{equation}
\begin{split} \label{qualit17a}
j_{1} = & 2 e \!\!\int\!\frac{d \varphi d \varphi'}{ (2\pi)^2 }
\int\!d x\!
\int\!d\vare \nu(\vare,x)\Gamma^{1}_{\varphi \rightarrow \varphi'}
\\
& \times\left[ f(\vare,x) - f(\vare,x + \vec{e}_x \Delta
\vec{R}_{\varphi \rightarrow \varphi'}) \right]
\end{split}
\end{equation}
describes the current in the absence of processes changing electron
energy by absorption or emission of a microwave field quantum with
energy $\omega$. This term contains virtual processes of electron
scattering in microwave field, which modify momentum scattering rate
off disorder. Thus, function $\Gamma^{1}_{\varphi \rightarrow
\varphi'} $ is the disorder scattering rate for the direction of
electron momentum from $\vec{n}_{\varphi}$ to $\vec{n}_{\varphi'}$
and has the following form within the Born approximation:
\begin{align}
\label{qualit28}
\Gamma^{1}_{\varphi \rightarrow \varphi'} =
\left[ \frac{ 1 }{ \tau_{\varphi - \varphi' } }- \frac{ \Pmw }{
\bar{\tau}_{\varphi - \varphi' } } \right]
\frac{ \nu\left(\vare, x + \Delta X_{\phi \rightarrow \phi'} \right)
}{ \nu_0 }\, ,
\end{align}
where $1 / \tau_{\varphi - \varphi' }$ is the disorder scattering
rate in the absence of electric and magnetic fields, and
\begin{align} \label{qualit37}
\frac{ 1 }{ \bar{\tau}_{\varphi - \varphi'} } = \frac{ 1 -
\cos(\varphi' - \varphi ) }{ \tau_{\varphi - \varphi'} }.
\end{align}
The dimensionless parameter
\begin{equation}
\Pmw = \frac{ v_F^2 e^2 E_{\omega }^2 }{ \omega^2 \left(\omega \mp
\wc \right)^2 } \label{Pmw=}
\end{equation}
is proportional to the microwave power. The upper (lower) sign in
Eq.~\eqref{Pmw=} corresponds to the right (left) circular
polarization of the incoming microwave radiation. Close to the
cyclotron resonance the dynamical screening of the microwave field
by electron system has to be taken into account. This screening
results in a modified form of parameter $\Pmw$, see
\oncite{dmitriev_JETP85}. The drift of the guiding centers due to
the microwave field effectively smears the disorder potential felt
by the electron. For that reason the scattering rate
$\Gamma^{1}_{\varphi \rightarrow \varphi'}$, Eq.~\eqref{qualit28},
is suppressed by the microwave radiation.
This scattering rate suppression is reminiscent of the ``motion
narrowing'' phenomenon and is further discussed in
Sec.~\ref{sec:Kin-Eq-Bilinear}.

Apart from the virtual processes of electron interaction with microwave field,
there are real processes, which are  accompanied by the absorption
and emission of photons.
The second term in \req{qualit15} takes into account the contribution to the electric current
from such processes:
\begin{align} \label{qualit17b}
j_{2} = & 2 e \sum_{\pm} \!\!\int\!\frac{d \varphi d \varphi'}{
(2\pi)^2 }
\int\!d x\!
\int\!d\vare \nu(\vare,x) \Gamma^{\pm}_{\varphi \rightarrow \varphi'}
\notag \\
& \times
\left[ f(\vare,x) - f(\vare \pm \omega,x + \vec{e}_x\Delta
\vec{R}_{\varphi \rightarrow \varphi'}) \right] \, .
\end{align}
The rate of such scattering processes can be
written in the form
\begin{align} \label{qualit35}
\Gamma^{\pm}_{\varphi \rightarrow \varphi'} = \Pmw \frac{ \nu\left(\vare
\pm \omega, x + \Delta X_{\varphi \rightarrow \varphi'} \right) }{
\nu_0 \bar{\tau}_{\varphi - \varphi' } }\, .
\end{align}

We emphasize that in the present subsection we do not consider multi-photon processes.
This approximation is
justified if the dimensionless parameter $\Pmw \ll 1$.
We analyze the case of the arbitrary parameter
$\Pmw$ in Sec.~\ref{sec:Arbitr-Power}, see also Sec.~\ref{sec:arbpower2}.

The angular integrals in \reqs{qualit17a} and \rref{qualit17b}
are restricted by the
conditions $\Delta X_{\varphi \rightarrow \varphi'}= \vec{e}_x
\Delta \vec{R}_{\varphi \rightarrow \varphi'} > 0$, and spatial
integration is limited to the stripe $- \Delta X_{\varphi
\rightarrow \varphi'}< x < 0 $. The electron density of states has
spacial dependence in constant electric fields. This dependence
appears under the spacial integral and we
discuss this dependence in more detail.
In a perpendicular magnetic field and in the absence of electric
fields, the density of states of a 2DES, $\nu(\vare)$, has an
energy modulation with period $\wc$
\begin{subequations}\label{qualit21}
\begin{align} \label{qualit21b}
\nu(\vare) = \nu_0\!\left(\!1 - 2 \lambda\cos \frac{ 2 \pi \vare }{
\wc } \!\right)\, ,
\end{align}
where $\nu_0 $ is the density of states in the absence of fields,
and the factor $\lambda = \exp( - \pi / \wc \tauq ) \lesssim 1$.
A constant electric field tilts electron density of states along its
direction, $\vec{e}_x$, resulting in spacial dependence of the
density of states:
\begin{align} \label{qualit21a}
\nu(\vare,x)= \nu(\vare + e E x )\, .
\end{align}
\end{subequations}

We present the results for the non-linear current in
strong dc electric fields, $\edc=2 \left| e \right| E\Rc/\wc \gg 1$,
where the differential resistance exhibits an oscillatory behavior.
Rigorous analysis of Sec.~\ref{sec:ACDC-Model} shows that in this
case the dominant contribution to the non-linear current originates
from the smooth component of the distribution function, which can be
taken as the Fermi-Dirac distribution function at temperature $T$:
\begin{equation} \label{ft=}
f_T(\vare)=\frac{1}{e^{\vare/T}+1}.
\end{equation}

Substitution of \req{qualit28} to \req{qualit17a}
gives the contribution $j_{1}$ to the current in the form
.
\begin{align} \label{qualit41}
j_{1} = & \frac{2 e}{ \nu_0 }\!\!\int\!\frac{d \varphi d
\varphi'}{ (2\pi)^2 }
\!\!\int\!\!d\vare
%\notag \\
%
%& \times
\nu\! \left(\vare \right)\!
\nu\!\left(\vare\!+\! e E \Delta X_{\varphi \rightarrow \varphi'}
\right)
\notag \\
& \times \Delta X_{\varphi \rightarrow \varphi'}
\left[ \frac{ 1 }{ \tau_{\varphi - \varphi' } } - \frac{ \Pmw }{
\bar{\tau}_{\varphi - \varphi'} } \right]
\notag \\
& \times\left[ f_T(\vare) - f_T(\vare + e E \Delta X_{\varphi
\rightarrow \varphi'}) \right] \, .
\end{align}
From \req{qualit41} we obtain
\begin{equation} \label{qualit43}
j_{1}=\sigma_{\rm D}E + \delta j_{1}^{(2)}.
\end{equation}
Here the first term is the zeroth order in $\lambda$, neglecting the
oscillations of electron density of states in magnetic fields in
Eq.~\eqref{qualit41}. This term corresponds to the classical Drude
contribution to the current at large Hall angles,
\begin{align} \label{qualit45}
\sigma_{\rm D} = e^2 \nu_0 \frac{ R_c^2 }{ \tau_{\rm tr} },\quad
\frac{1}{\tau_{\rm tr}}=\int\limits_{-\pi}^{+\pi}
\frac{d\theta}{2\pi}\frac{1-\cos\theta}{\tau_\theta} ,
\end{align}
where $\tau_{\rm tr}$ is the transport scattering time. The second
term in \req{qualit43} represents a contribution to the current,
which is non-linear in the applied electric field $E$ and quadratic
in parameter $\lambda$:
\begin{align} \label{qualit48}
\delta j_{1}^{(2)}\! = &\! 2\lambda^2 e^2\nu_0 E
\!\!\int\!\frac{d \varphi d \varphi'}{ (2\pi)^2 }
\! \left[ \frac{ 1 }{ \tau_{\varphi - \varphi' } } - \frac{ \Pmw }{
\bar{\tau}_{\varphi - \varphi'} } \right]
\notag \\
& \times \left[ \Delta X_{\varphi \rightarrow \varphi'} \right]^2
\!\cos \!\frac{ 2 \pi e E \Delta X_{\varphi \rightarrow \varphi'} }{
\wc }\, .
\end{align}
The contribution of the first order in $\lambda$, which is omitted
in \req{qualit43}, describes the Shubnikov-de Haas oscillations and
can be estimated as $\propto \sigma_D \lambda_T \lambda$. Here the
additional small prefactor $\lambda_T = \exp( - 2 \pi^2 T/ \wc)$
appears as a result of averaging of the rapid oscillations in the
density of states over thermal energies $|\vare|\lesssim T$. The
contribution of the second order in $\lambda$ contains a square of
the oscillating component of the density of states and is not
exponentially suppressed after integration over thermal energy
window.  In this paper we consider the limit of relatively high
temperatures $T \gg \wc/2\pi$. The latter condition is normally
satisfied in experiments performed at the non-linear regime in dc
electric field. Under this condition the quadratic in $\lambda$
contribution dominates over the Shubnikov-de Haas contribution
linear in $\lambda$.

For $\edc \gg 1 $ the angular integrations in Eq.~\eqref{qualit48}
can be performed in the stationary phase approximation. The main
contribution comes from the scattering within narrow intervals
centered at $\varphi = \pm \pi/2$, $\varphi' = \mp \pi/2$. These
back scattering processes correspond to the $2\Rc$ jumps of electron
guiding center along the electric field. Equation \eqref{qualit48}
yields~\cite{VAG}
\begin{align} \label{qualit51}
\delta j_{1}^{(2)} \approx (2 \lambda)^2 \left( 1 - 2 \Pmw
\right) \frac{ e \nu_0 \vF }{\pi^2 \tau_{\pi}}
\sin 2 \pi \edc \, .
\end{align}

A similar analysis of the contribution \req{qualit17b} to the
electric current due to the real processes of absorption and emission
of microwave photons gives:
\begin{align} \label{qualit56}
j_{2} \approx & (2 \lambda)^2 \Pmw \frac{ e \nu_0 \vF }{\pi^2
\tau_{\pi}}
\notag \\
& \times \left[ \frac{ \edc + \eac} { \edc } \sin 2 \pi (\edc +
\eac)
\right. \notag \\
& \left. + \frac{ \edc - \eac} { \edc } \sin 2 \pi (\edc - \eac)\!
\right] \, .
\end{align}
We emphasize that both \reqs{qualit51} and \rref{qualit56} are
expressed through the parameter $\edc \propto E R_c$. These two
expressions are obtained within a stationary phase approximation and
correspond to processes representing shifts of electron cyclotron
trajectories by distance $2\Rc$ along the applied dc electric field.
Therefore, \reqs{qualit51} and \rref{qualit56} have a simple
geometrical interpretation in terms of commensurability between the
space modulation of electron density of states by electric field and
the maximal displacement $2\Rc$ in a single scattering process off
disorder. The expression for the non-linear current beyond the
saddle point approximation is evaluated in
Sec.~\ref{sec:ACDC-NL-Curr}.

In general, we represent the total current as a sum of the linear
Drude term $\sigma_{\rm D}E$ and the non-linear term $\delta j$,
which arises due to oscillatory density of states in perpendicular
magnetic field: \be j=\sigma_{\rm D}E+\delta j. \label{current=} \ee
For the overlapping Landau levels, when the density of states is
given by \req{qualit21b}, the non-linear contribution is quadratic
in parameter $\lambda$.

Within a bilinear response to the applied microwave electric field
and in strong dc electric fields we combine Eqs.~\eqref{qualit15}
and \eqref{qualit43} and obtain
\begin{align} \label{qualit59}
\delta j = \delta j_{1}^{(2)} + j_{2}.
\end{align}
Equations~\eqref{qualit51}, \eqref{qualit56} and \eqref{qualit59}
represent some of the main results of this paper.
As we discuss in Sec.~\ref{sec:ACDC}, despite its simplicity,
Eq.~\eqref{qualit56} explains pronounced features of the
transport measurement reported in Refs.~\onlinecite{zudovACDC07,zudovACDC08}.

\subsection{Current at arbitrary microwave powers in strong dc electric
field}
\label{sec:arbpower2}

The result of \req{qualit59} is just a limiting case of expression
obtained in Sec.~\ref{sec:Arbitr-Power} for the current in strong dc
electric fields. In this case we again can neglect the modification
of the electron distribution function by electric fields and
consider only the displacement mechanism for generation of the
non-linear component $\delta j$ of the current \req{current=}. We
found the following expression
\begin{equation}
\begin{split} \label{ArbitrPower=}
\delta j =& (2\lambda)^2 \frac{ |e|\nu_0 \vF }{ \pi^2\tau_{\pi} }
\Big[
\sin 2 \pi \edc J_0\left(4 \sqrt{\Pmw } \sin \pi \eac
\right)\\
&+ \frac{ 2 \eac }{ \edc } \cos 2 \pi \edc \cos \pi \eac
\sqrt{\Pmw } J_1\left(4 \sqrt{\Pmw } \sin \pi \eac \right) \Big]\, .
\end{split}
\end{equation}
Here $J_n(x)$ are the Bessel functions. Performing an expansion in
Eq.~\eqref{ArbitrPower=} to the first order in $\Pmw$, we recover
\req{qualit59} in terms of
Eqs.~\eqref{qualit51} and \eqref{qualit56}.

\subsection{Weak electric fields}
\label{sec:Qualit-Other}

Above results were obtained under assumption that the distribution
function of electrons is given by the Fermi distribution function,
see \req{ft=}. In the presence of electric fields the distribution
function deviates from the equilibrium
configuration. This non-equilibrium distribution
function affects significantly the electric current in sufficiently
weak electric fields.
For smooth disorder, we recover the result of \oncite{DVAMP} for
the non-linear  contribution $\delta
j$ to current in terms of dimensionless parameters
$\edc$, $\eac$, defined by \req{eac_edc}, and $\Pmw$, defined by
\req{Pmw=}:
\begin{equation}
\label{qualit66} \delta j = 2 \lambda^2  \sigma_{\mathrm{D}} E
%%
% \\
%%
%& \times
\frac{ \pi^2\edc^2
+ 2\pi \eac \Pmw  \sin 2 \pi \eac }
{ \tautr/ \tauin + \pi^2 \edc^2 /2
+ 2 \Pmw \sin^2\pi \eac } \, .
%
%& \times \left[ 1 - 2 \Pmw (\tautr/ \tau_{tr,2}) \sin^2\pi \eac
%\right]\, ,
\end{equation}
Here $1/\tauin$ is the relaxation rate of the non-equilibrium
component of the electron distribution function due to
electron-electron interaction. In the presence of sharp disorder,
the ``displacement'' contribution, arising due to the modification
of electron scattering rate off disorder may become comparable to
the ``inelastic'' contribution \req{qualit66}. The latter
contribution survives only in relatively weak electric fields,
$\edc\lesssim 1$. In stronger electric fields, $\edc\gg 1$, it
contains an extra small factor $\tauq/(\tau_\pi \edc)$ and the
``displacement'' mechanism becomes more important, see
Sec.~\ref{sec:ACDC-Model}.

\section{Quantum Kinetic equation}
\label{sec:Kin-Eq}
In the present section, we study a disordered 2DEG subject to
in-plane electric fields and derive the quantum kinetic equation,
following~\oncite{VA03}, but consider a mixed disorder characterized
by scattering amplitude which is finite for an arbitrary scattering
angle. The Dyson equation for the disorder averaged electron Green
function
\begin{align} \label{KinEq13}
\left(i \partial_t - \hat{H} \right)\widehat{G}(tt') = \widehat
1\delta(t-t') +\int dt_1 \widehat{\Sigma}(tt_1) \widehat{G}(t_1t')
\, ,
\end{align}
where $\hat H$ is the one-electron Hamiltonian in the absence of
disorder. Both the Green function $\widehat{G}$ and the self-energy
$\widehat{\Sigma}$ for electron scattering off disorder are matrices
in the Keldysh space:
\begin{equation}
\widehat{G}=\left(\begin{array}{cc}
\hat G^R & \hat G^K \\
0 & \hat G^A
\end{array}\right),\quad
\widehat{\Sigma}=\left(\begin{array}{cc}
\hat \Sigma^R & \hat \Sigma^K \\
0 & \hat \Sigma^A
\end{array}\right).
\end{equation}
Matrix $\widehat 1$ stands for unit matrix in the Keldysh space as
well as in the one-electron Hilbert space.

Below we assume that the disorder potential is a combination of the
long-range potential with correlation length $\xi$ created by remote
positively charged donors, and the short-range potential with much
smaller correlation length. When  conditions $\xi \ll \lambda_H$
and $\pF l \gg 1$ are satisfied, the self consistent Born
approximation for the self-energy calculation is
applicable.~\cite{RaikhShahbazyan1993} In this case, the
expression for the self-energy takes the form~\cite{VA03}
\begin{align} \label{KinEq31}
\widehat{\Sigma}(\hat{\vec{p}}) = \int \frac{ d^2 \vec{q} }{ (2
\pi)^2 } W(|\vec{q}|)
\left[e^{i \vec{q} \hat{\vec{r}}} \widehat{G}(\hat{\vec{p}}) e^{-i
\vec{q} \hat{\vec{r}}} \right]\, .
\end{align}
Here the function $W(|\vec{q}|)$ is introduced as a Fourier
transform of the correlation function of a Gaussian disorder
potential
\begin{align} \label{KinEq29}
\langle U(\vec{r}_1) U(\vec{r}_2) \rangle =
\int \frac{ d^2 q }{ (2 \pi)^2 } W(q) e^{ i \vec{q} (\vec{r}_1 -
\vec{r}_2) } \, .
\end{align}

The external electric field can be eliminated by transferring to the
moving reference frame, $\vec{r }\rightarrow \vec{r} +
\vec{\zeta}(t)$, where $\vec{\zeta}(t)$ describes a two-dimensional
electron motion in crossed electric and magnetic fields:
\begin{align} \label{KinEq39}
\partial_t \vec{\zeta}(t) = \left( \frac{\partial_t - \wc
\hat{\epsilon} }{ \partial^2_t + \wc^2 } \right) \frac{ e \vec{E}(t) }{
m }\, .
\end{align}
Here $\wc = \left| e \right| B / m c $ is the cyclotron frequency,
and $\hat{\epsilon}$ is the antisymmetric tensor: $\hat{\epsilon} \vec{E}
= \vec{E} \times \vec{e}_z $ for any vector $\vec{E}$ lying in the
$x-y$ plane of the 2DES, see Fig.~\ref{fig:shift}.
In the moving reference frame, the disorder potential acting on
electrons is time dependent, and \req{KinEq31} becomes
\begin{align} \label{KinEq43}
\widehat{\Sigma}(\hat{\vec{p}}) = \int \frac{ d^2 q }{ (2 \pi)^2 }
W_{t_1 t_2}( \vec{q} ) [e^{i \vec{q} \hat{\vec{r}}} \widehat{G}(
\hat{\vec{p}} ) e^{-i \vec{q} \hat{\vec{r}}}],
\end{align}
where
\begin{subequations}\label{KinEq46}
\begin{align} \label{KinEq46a}
W_{t_1 t_2}(\vec{q}) = W(q)e^{ i \vec{q} \vec{\zeta}_{t_1 t_2} },
\end{align}
\begin{align} \label{KinEq46b}
\vec{\zeta}_{t_1 t_2} = \vec{\zeta}(t_1) - \vec{\zeta}(t_2).
\end{align}
\end{subequations}

To proceed further we introduce the operator of the guiding center coordinate
$\hat{\vec{R}}$:
\begin{align} \label{KinEq51}
\hat{H}= \frac{ \hat{\vec{p}}^2 }{ 2 m } - \mu, \quad
\hat{\vec{r}} = \hat{\vec{R}} + \lambda_H^2 \hat{\epsilon}
\hat{\vec{p}} \, ,
\end{align}
where $\lambda_H = (c \hbar / \left| e \right| B )^{1/2} $ is the
magnetic length.
The commutation relations between the operators of the guiding center $\hat{\vec{R}}$
and momentum $\hat{\vec{p}} $ are:
\begin{align} \label{KinEq56}
[\hat{R}_{\alpha}, \hat{R}_{\beta}] = i \lambda_H^2
\vare_{\alpha\beta}, \quad [\hat{p}_{\alpha}, \hat{p}_{\beta}]
=-\frac{ i}{ \lambda_H^2 } \vare_{\alpha\beta}, \quad
[\hat{\vec{R}}, \hat{\vec{p}}] = 0 \, .
\end{align}
The equation \eqref{KinEq43} takes the form
\begin{align} \label{KinEq59}
\widehat{\Sigma}(\hat{\vec{p}}) = \int \frac{ d^2 q }{ (2 \pi)^2 }
W_{12}( \vec{q} ) [e^{i \vec{q} \hat{\epsilon} \hat{\vec{p}}
\lambda_H^2 } \widehat{G}( \hat{\vec{p}} ) e^{ - i \vec{q}
\hat{\epsilon} \hat{\vec{p}} \lambda_H^2 }] \, .
\end{align}

We first analyze the retarded and advanced components of the Dyson
equation, \req{KinEq13}, which determine electron spectrum. Then, we
reduce the equation for the Keldysh component of  \req{KinEq13} to
the kinetic equation for the electron distribution function.

\subsection{Electron spectrum}
\label{sec:Kin-Eq-Spectrum}

The Dyson equation, \req{KinEq13}, for the retarded component of
Green function is given by
\begin{align} \label{Spectrum12}
& \left[ i \partial_t - \hat{H} \right]
\hat{G}^R(t,t_1 ; \hat{ \vec{p} } )\notag \\
& \phantom{pp} =\frac{ \delta(t - t_1) }{ 2 \pi } + \int_{t_1}^{t} d
t_2 \hat{\Sigma}^R(t,t_2) \hat{G}^R(t_2,t_1; \hat{\vec{p}})
\end{align}
along with the self consistency equation \eqref{KinEq59}. We limit
the analysis to the first order in the parameter $\lambda = e^{- \pi
/ \wc \tauq }$. The zeroth order solution in $\lambda$ corresponds
to the solution in the absence of a magnetic field. The standard
answer for the self-energy in this case is
\begin{align} \label{Spectrum15}
\hat{\Sigma}^R_0(t,t') = - \frac{ i }{ 2 \tauq } \hat{I}_{\rm e}
\delta(t-t') \,,
\end{align}
where $\hat{I}_{\rm e}$ is the unity operator in the coordinate space.

The
Green function corresponding to \eqref{Spectrum15} is obtained by
solving \eqref{Spectrum12}
\begin{align} \label{Spectrum19}
\hat{G}^R_0(t,t') = - i \theta( t- t') e^{-i \hat{H}(t - t') }
e^{-(t-t') / 2 \tauq } \, .
\end{align}
We consider the first iteration of the self consistent Born approximation.
For this purpose, we
substitute \req{Spectrum19} to the retarded matrix element of
\req{KinEq59}:
\begin{align} \label{Spectrum21}
\hat{\Sigma}'^{R}_{tt'}(\vec{p}) = &- i \theta( t- t') e^{-(t-t') /
2 \tauq
}\int \frac{ d^2 q }{ (2 \pi)^2 } W_{tt'}(\vec{q}) \notag \\
& \times e^{i \vec{q} \vare \hat{\vec{p}} \lambda_H^2 } e^{-i
\hat{H}(t - t') } e^{ - i \vec{q} \vare \hat{\vec{p}} \lambda_H^2 }
\, .
\end{align}
We notice that the operator product in the integrand in
\eqref{Spectrum21} can be written as
\begin{align} \label{Spectrum23}
e^{i \vec{q}\hat{\epsilon} \hat{\vec{p}} \lambda_H^2 } e^{-i \hat{H}  t
} e^{ - i \vec{q} \hat{\epsilon} \hat{\vec{p}} \lambda_H^2 } = e^{-i
\hat{H}
t } e^{i \vec{q} \hat{\epsilon} \hat{\vec{p}}_{  t} \lambda_H^2 } e^{ -
i \vec{q} \hat{\epsilon} \hat{\vec{p}} \lambda_H^2 },
\end{align}
where the time dependent operators are  $\hat{\vec{p}}_t = e^{i \hat{H} t
}\hat{\vec{p}} e^{-i \hat{H} t}$. To find the time dependence of the
operators, it is convenient to introduce~\cite{ChiralRes}
``raising,'' $\hat{p}^{+}$, and ``lowering,'' $\hat{p}^{-}$,
operators, defined as $\hat{p}^{\pm} = (p_x \pm i p_y) / \sqrt{ 2 }
$ and obeying the commutation relation $ [\hat{p}^{+}, \hat{p}^{-}]
= -1/ \lambda_H^2 $. The introduced notations for $\hat{p}^{\pm}$
are useful as new operators have simple form in the interaction
picture
\begin{align} \label{Spectrum24}
\hat{p}^{\pm}_t = e^{\pm i \wc t}\hat{p}^{\pm} .
\end{align}
We also introduce the notations for vectors ${q}^{\pm}=(q_x\pm i
q_y)/\sqrt{2}$ and write $\vec{q}\vare \hat{\vec{p}} = {q}^{-}
\hat{{p}}^{+} - {q}^{+} \hat{{p}}^{-}$. Using \req{Spectrum24} and
the operator relation $e^{\hat A} e^{\hat B} = e^{\hat A+\hat
B}e^{(1/2)[\hat A,\hat B]}$, we further transform \req{Spectrum23}
to
\begin{align} \label{Spectrum27}
& e^{i \vec{q} \vare \hat{\vec{p}}_{ t} \lambda_H^2 } e^{ - i
\vec{q} \vare \hat{\vec{p}} \lambda_H^2 }
\notag \\
& = \exp\left\{ 2 i \lambda_H^2 (\tilde{{q}}^-_{ t}
\hat{{p}}^+ + \tilde{{q}}^+_{ t} \hat{{p}}^- )
\sin\left[\frac{ \wc t}{ 2 }   \right] \right\} \notag \\
& \times\exp\left\{ - i \lambda_H^2 \frac{ q^2 }{ 2 } \sin\left[ \wc
t \right]\right\}\, .
\end{align}
In equation \eqref{Spectrum27}, we denoted ${\tilde{q}}_t^{\pm} =
{q}^{\pm} e^{\pm i \wc t / 2 }$, or in other words, the vector
$\tilde{\vec{q}}_t$ is obtained from the vector $\vec{q}$ by a rotation
on angle $ \wc t / 2 $.
Using Eqs. \eqref{Spectrum23} and \eqref{Spectrum27} we can write
Eq.~\eqref{Spectrum21} in the following form
\begin{align} \label{Spectrum29}
\hat{\Sigma}'^{R}_{tt'}(\vec{p}) & = - i \theta( t- t') e^{-
(i\hat{H}+1/2 \tauq) (t - t') } \int \frac{ d^2 \vec{q} }{ (2 \pi)^2
}
\notag \\
&\times W_{tt'}(|\vec{q}|) \exp\left\{ 2 i \lambda_H^2
\tilde{\vec{q}}_{t-t'} \hat{\vec{p}}
\sin\left[\frac{ \wc }{ 2 } (t - t') \right] \right\} \notag \\
& \times\exp\left\{ - i \lambda_H^2 \frac{ q^2 }{ 2 } \sin\left[ \wc
(t - t') \right] \right\} \, .
\end{align}

It is illustrative to consider the zero field limit. In this case
the momentum operator becomes a $c$-number and \req{Spectrum29}
yields
\begin{align} \label{Spectrum31}
\hat{\Sigma}'^{R}_{tt'}(\vec{p}) = & - i \theta( t- t')e^{-(t-t') /
2 \tauq }
\nonumber\\
&\times\int \frac{ d^2 q }{ (2 \pi)^2 } W(q)
e^{- i \xi_{\vec{p} - \vec{q}}(t - t') } \, .
\end{align}
Changing integration variables in \eqref{Spectrum31} to
$\xi_{\vec{p} - \vec{q}}$ and the angle formed by the vector
$\vec{p} - \vec{q}$ with some fixed direction, we show the
consistency of the employed approximation, namely
\begin{align} \label{Spectrum35}
\hat{\Sigma}'^{R}(\vec{p}) = \hat{\Sigma}_0^R \,
\end{align}
with $\hat{\Sigma}_0^R$ defined by \req{Spectrum15} and
\begin{align} \label{Spectrum38a}
\frac{ 1 }{ \tauq } = \int \frac{ d \theta }{ 2 \pi } \frac{ 1 }{
\tau_{\theta} }.
\end{align}
The scattering rate
\begin{align}
\label{Spectrum38}
\quad \frac{1 }{ \tau_{\theta} } = 2 \nu_0 \pi W \left( 2 \pF \sin
\frac{ \theta }{ 2 } \right) \, .
\end{align}
off disorder on angle $\theta$ can be written in terms of its angular harmonics
\be
\frac{1}{\tau_\theta}=\sum_{n=-\infty}^{+\infty}\frac{e^{in\theta}}{\tau_n},\quad
\tau_{-n}=\tau_n.
\label{taundef=}
\ee
Then, $\tauq=\tau_0$.

In finite magnetic fields, we notice that the exponential factors in
\req{Spectrum29} are $2 \pi /\wc$ periodic. The argument of these
exponents vanishes at $t-t'= l\Tc$ with integer $l$ and the
integral in \req{Spectrum29} diverges; $\Tc=2\pi/\wc$ is the cyclotron period. We argue that this integral
gives, in fact, a $\delta$-peak of the width $\delta t \approx 1 /
E_{\rm F} $ for time difference $t-t'= l \Tc$. Indeed, this
statement is obvious if the operators in the exponent of
\req{Spectrum29} can be treated as commuting. For time intervals $|
t - t' -  l \Tc | \lesssim 1 /\sqrt{ E_{\rm F} \wc }$ the
commutator of the two operators in the exponent of \req{Spectrum29}
is small since, in this case, each operator is multiplied by
$\sin(\wc(t-t')/2)\lesssim \sqrt{\wc/E_{\rm F}}$ and we can apply the
same argument as the one used in zero magnetic field. On the other
hand, for $ 1 /\sqrt{ E_{\rm F} \wc }
\lesssim | t - t' -  l \Tc | \lesssim  \Tc/2$ the result of the integration in
\eqref{Spectrum29} vanishes because of the rapid oscillations of the
exponent. This can be checked explicitly by calculation of the
matrix elements of the self energy \req{Spectrum29}.
We conclude that the non-commutativity of the operators can be
ignored in Eq.~\eqref{Spectrum29} and we can apply our zero field
considerations whenever $ t - t' $ is a multiple of the cyclotorn period $ \Tc $.

The contributions to the self-energy \eqref{Spectrum21} with $l>1$
are proportional to $\lambda^l$ and, therefore, can be neglected in
moderately weak magnetic fields, when $\lambda\ll 1$ . We stress
that the terms with $l > 1$ are beyond the accuracy of the first
iteration of the self-consistent scheme. The corrections of the
higher orders in $\lambda$ can be taken into account in the spirit
of~\oncite{VA03}, where the small scattering angle was only
considered. In this paper, we restrict our analysis to terms
$l=0,1$, sufficient in not too strong magnetic fields.

%Once the small quantum contribution ($\propto \lambda$) has been
%singled out, the remaining variables in Eq.~\eqref{Spectrum29} can be
%treated as commuting.

Introducing a new variable $\vec{p}' = \vec{p} - \vec{q}$ in
\eqref{Spectrum29} and neglecting the variation of the function
$W_{t,t'}(\vec{p}' - \vec{p})$ given by \req{KinEq46} (see bellow)
we can easily perform the integration over the absolute value
$|\vec{p}'|$. This integration results in
\begin{align} \label{Spectrum44}
\hat{\Sigma}'^{R}_{t,t';\varphi} = & \frac{1}{i} \left( \frac{1}{2}
\delta(t-t') - \lambda \delta(t - t' - \Tc ) \right)
%\notag \\
%
%& \times\!
\mathcal{\hat{K}}_{t,t';\varphi}\{1\},
\end{align}
where we have defined the integral kernel
\begin{align} \label{Spectrum49}
\mathcal{\hat{K}}_{t,t';\varphi}\{F(\varphi)\} =
\int \frac{ d \varphi' }{ 2 \pi
}
\frac{ e^{ i p_{\rm F}
(\vec{n}_{\varphi}-\vec{n}_{\varphi'})\vec{\zeta}_{t,t' } }}{
\tau_{\varphi - \varphi'} } F(\varphi')\,
\end{align}
with an arbitrary function $F(\varphi)$; $\vec{n}_{\varphi}=(\cos\varphi;
\sin\varphi;0)$.
The negative sign of the second term in Eq.~\eqref{Spectrum44}
corresponds to the chemical potential $\mu = s \omega_c$, with
integer $s$, such that the density of states $\nu(\mu)$ is at
minimum. We note that the exact position of the chemical potential
is of no importance for our final results.

Equation \eqref{Spectrum44} is valid if the variation of the matrix
element for the transitions in the moving reference frame,
$W_{t,t'}(\vec{p}' - \vec{p})$ can be neglected in the course of
integration over the absolute value $|\vec{p}'|$. This requirement
leads to the limitation on the strength of electric fields.
Since the relevant time scale is $|t - t' -  l \Tc | \lesssim
\tauq \ll \Tc $, which insures that the parameter $\lambda$ can
be defined in Eq.~\eqref{Spectrum44} unambiguously, the important
range of the integration is $|p'| < 1 / v_{\rm F} \tauq$.
The requirement of smoothness of $W_{t,t'}(\vec{p}' - \vec{p})$ in
the momentum variable means that the distance electron drifts over
one cyclotron period must be smaller than the quantum length $
v_{\rm F} \tauq $. In the case of a constant electric field, this
condition amounts to $v_D / \wc \ll v_{\rm F} \tauq $, or
equivalently, $\edc \ll E_{\rm F} \tauq$. If this condition is not
met, the amplitude of oscillations as a function of $\edc$ is
reduced because of the finite broadening in time of the self-energy.
In experiments,\cite{zudovDC06,zudovACDC07} $E_{\rm F} \tauq \approx
10^2$, $\edc \lesssim 5$ and the above condition is satisfied.
However, the amplitude of oscillations in \oncite{zudovDC06} shows
tendency to decrease when $\edc$ increases. In our model, this
tendency can be accounted for by taking into account the dependence of the function
$W_{t,t'}(\vec{p}' - \vec{p})$ on the absolute value of momenta,
which becomes stronger as $\edc$ increases.

\subsection{Electron distribution function}
\label{sec:Kin-Eq-Distrib-Func}

In the previous subsection, we analyzed the effect of disorder on
the spectral characteristics of electron Green function, determined
by the retarded and advanced components of the Green function. Now
we reduce the equation for the Keldysh component of the Green
function to the kinetic equation for the electron distribution
function $\hat f$, related to the Keldysh component $\hat G^K$
through the standard expression
\begin{align} \label{KinEq18}
\hat{G}^K = \hat{G}^R - \hat{G}^A - 2 [\hat{G}^R \hat{f} - \hat{f}
\hat{G}^A] \, .
\end{align}
In the present analysis, the Wigner transformation $f(t,t';\vec{R},\vec{p})$
of the distribution function $\hat f$ in time and coordinate variables
is independent of the ``center of mass''
coordinate $\vec{R} = (\vec{r} + \vec{r}')/2 $ due to the
translational symmetry. The peaked structure of the retarded
and advanced Green functions in Eq.~\eqref{KinEq18} makes
$f(t,\epsilon;\vec{R},\vec{p})$ to be independent of the absolute value of the
momentum, $\vec{p} $. The dependence on the direction
of the momentum is still to be retained. %%
Notice that although the two components of the momentum operator are
not commuting, the momentum direction is well defined in the
quasi-classical regime $E_{\rm F} \gg \wc$. Therefore, we can write
$f(t,t';\vec{R},\vec{p})=f(t,t';\vec{n}_\varphi)= f_{t,t';\varphi}$.
The solution of the resulting quantum kinetic equation presented in
Sec.~\ref{sec:ACDC} is consistent with the assumptions made above.

The distribution function $f(t,t';\vec{n}_\varphi)$ obeys the
following kinetic equation:
\begin{align} \label{KinEq21}
[(\partial_t + i \hat{H} );\hat{f}]=\mathrm{St}_{\mathrm{dis}} \{\hat{f} \}
+\mathrm{St}_{\rm ee} \{ \hat{f} \},
\,
\end{align}
where the notation $[\cdot;\cdot]$ stands for the commutator in one-particle
Hilbert space and the time variable.
In Eq.~\eqref{KinEq21} the collision integral $\mathrm{St}_{\mathrm{ee}} \{ f \}$
describes the electron-electron interaction and is discussed in the end of this section.
The collision integral $\mathrm{\widehat{St}}_{\mathrm{dis}} \{\hat{f} \}$ represents
scattering off disorder and
can be written as the sum of scattering "out" and "in" terms
\begin{align} \label{KinEq23}
\mathrm{St}_{ \mathrm{dis} } = \mathrm{St}_{ \mathrm{out} } + \mathrm{St}_{
\mathrm{in} } \, ,
\end{align}
where
\begin{subequations}
\begin{align} \label{KinEq25a}
i \mathrm{St}_{\mathrm{out} } \{\hat{f}\} & =
[\hat{\Sigma}^R \hat{f }- \hat{f} \hat{\Sigma}^A ], \\
\label{KinEq25b}
i \mathrm{St}_{\mathrm{in} } \{\hat{f}\} & = \frac{ 1 }{ 2
}[\hat{\Sigma}^K - \hat{\Sigma}^R + \hat{\Sigma}^A ] \, .
\end{align}
\end{subequations}
Next, we analyze the collision integral in Eq.~\eqref{KinEq21}.
We start our analysis with the scattering-"out" term,
Eq.~\eqref{KinEq25a}, written as
\begin{equation} \label{Collision13}
i\mathrm{St}_{\mathrm{out} } \{ \hat{f} \}_{t,t';\varphi} =
\int d t''
\left[ \hat{\Sigma}'^R_{t,t'';\varphi} f_{t'',t';\varphi}
%\right. \notag \\
%%
%& \left.
- f_{t,t'';\varphi} \hat{\Sigma}'^A_{t'',t';\varphi} \right]\, .
\end{equation}
Using Eq.~\eqref{Spectrum44} we can perform integration over
intermediate time $t''$ in \req{Collision13} and obtain
\begin{align} \label{Collision15}
& \mathrm{St}_{\mathrm{out} } \{ \hat{f} \}_{t,t';\varphi}  =
- \mathcal{\hat{K}}_{t,t;\varphi}\{1\} f_{t,t';\varphi}
\notag \\
+& \lambda \left(\mathcal{\hat{K}}_{t,t -\Tc;\varphi}\{1\} f_{t- \Tc,t';\varphi} %\notag \\
+ f_{t, t' - \Tc;\varphi} \mathcal{\hat{K}}_{t' - \Tc,t';\varphi}\{1\} \right)\, ,
\end{align}
where the operator $\hat{\mathcal{K}}_{t,t';\varphi}\{1\}$  acts on
a unity as defined by \req{Spectrum49} and $\Tc=2\pi/\wc$.

We now turn to the consideration of the scattering-"in" term,
Eq.~\eqref{KinEq25b}. We express the self energies through the Green
functions, using the self consistency condition, Eq.~\eqref{KinEq43}
and the parametrization in Eq.~\eqref{KinEq59}
\begin{align} \label{Collision18}
\mathrm{St}_{\mathrm{in} } \{ \hat{f} \}_{t,t'} = &
i \int \frac{ d^2 \vec{q} }{ (2 \pi)^2 } W_{t,t'}(\vec{q}) \notag \\
& \times e^{i \vec{q }\vare \hat{\vec{p}} \lambda_H^2 } \left[
\hat{G}^R \hat f - \hat f \hat{G}^A \right] e^{ - i \vec{q} \vare
\hat{\vec{p}} \lambda_H^2 } \, ,
\end{align}
where the Green functions are given by \req{Spectrum19}. Commuting
the retarded (advanced) Green function with the left (right)
exponent, we rewrite \eqref{Collision18} as
\begin{align} \label{Collision21a}
& \mathrm{St}_{\mathrm{in} } \{ \hat{f} \}_{t,t'} = i
\int \frac{ d^2 \vec{q} }{ (2 \pi)^2 } \int d t'' W_{t,t'}(\vec{q}) \notag \\
& \times \left[
\hat{G}^R_{t,t''} e^{i \vec{q} \vare \hat{\vec{p}}_{t - t''}
\lambda_H^2 } f_{t'',t'}(\hat{\vec{p}}) e^{ - i \vec{q} \vare
\hat{\vec{p}} \lambda_H^2 }
\right. \notag \\
& \left. - e^{  i \vec{q} \vare \hat{\vec{p}} \lambda_H^2 }
f_{t,t''}(\hat{\vec{p}}) e^{ - i \vec{q} \vare \hat{\vec{p}}_{t' -
t''} \lambda_H^2 } \hat{G}^A_{t'',t'} \right] \, .
\end{align}
Using $e^{ i \vec{q} \vare
\hat{\vec{p}} \lambda_H^2 } f_{t,t'}(\hat{\vec{p}}) e^{ - i \vec{q} \vare
\hat{\vec{p}} \lambda_H^2 }= f_{t,t'}(\hat{\vec{p}}-\vec{q})$,
we commute the exponents with the distribution function:
\begin{align} \label{Collision21}
&\mathrm{St}_{\mathrm{in} } \{ \hat{f} \}_{t,t'} = i
\int \frac{ d^2 \vec{q} }{ (2 \pi)^2 } \int d t'' W_{t,t'}(\vec{q}) \notag \\
& \times \left[
\hat{G}^R_{t,t''} e^{i \vec{q} \vare \hat{\vec{p}}_{t - t''}
\lambda_H^2 }e^{ - i \vec{q} \vare \hat{\vec{p}} \lambda_H^2 }
f_{t'',t'}(\hat{\vec{p}} - \vec{q})
\right. \notag \\
& \left. - f_{t,t''}(\hat{\vec{p}} - \vec{q})e^{ i \vec{q} \vare
\hat{\vec{p}} \lambda_H^2 } e^{ - i \vec{q} \vare \hat{\vec{p}}_{t'
- t''} \lambda_H^2 } \hat{G}^A_{t'',t'} \right] \, .
\end{align}
Following the same line of arguments as in the derivation of
\eqref{Spectrum44} we put \eqref{Collision21} into the form
\begin{align} \label{Collision25}
& \mathrm{St}_{\mathrm{in} } \{ f  \}_{t,t';\varphi}=
\int d t'' \notag \\
& \times \left\{
\left( \frac{ 1 }{ 2 } \delta( t - t'') - \lambda \delta(t - t'' -
\Tc ) \right)
\mathcal{\hat{K}}_{t,t';\varphi} \{f_{t'',t';\varphi}\} \right. \notag \\
& \left. +
\left( \frac{ 1 }{ 2 } \delta( t' - t'') - \lambda \delta(t' - t'' -
\Tc ) \right) \mathcal{\hat{K}}_{t,t';\varphi}\{f_{t,t'';\varphi}\}
\right\},
\end{align}
where  $\mathcal{\hat{K}}_{t,t';\varphi} \{f_{t_1,t_2;\varphi}\}$ is
defined by \req{Spectrum49}.
Performing the time integration in \req{Collision25} we find
\begin{align} \label{Collision28}
\mathrm{St}_{\mathrm{in} }\{ \hat{f} \}_{t,t';\varphi} = &
\mathcal{\hat{K}}_{t,t';\varphi} \{ f_{t,t';\varphi}\}
- \lambda \mathcal{\hat{K}}_{t,t';\varphi} \{f_{t - \Tc
,t';\varphi}\}
\notag \\
& - \lambda \mathcal{\hat{K}}_{t,t';\varphi} \{f_{t ,t'-
\Tc;\varphi} \}\, .
\end{align}
Collecting  \reqs{Collision15} and \eqref{Collision28} and using
$i[\hat H; f_{t,t';\varphi}]=\wc\partial_\varphi f_{t,t';\varphi}$
we can finally write down the kinetic equation
\begin{align} \label{Collision32full}
\left[ \frac{ \partial }{ \partial t } + \frac{
\partial }{
\partial t' } + \wc \frac{ \partial }{ \partial \varphi }
\right] f_{t,t';\varphi} = \mathrm {St}_{\mathrm{dis}}\{ f \}_{t,t';\varphi}+\mathrm {St}_{\mathrm{ee}}\{ f \}_{t,t';\varphi}
\end{align}
with the collision integral
\begin{align} \label{Collision36}
\mathrm {St}_{\mathrm{dis}}& \{ f \}_{t,t';\varphi} = \mathcal{\hat{K}}_{t,t';\varphi} \{f_{t,t';\varphi}\} -
\mathcal{\hat{K}}_{t,t;\varphi}\{1\} f_{t,t';\varphi}
\notag \\
& - \lambda \mathcal{\hat{K}}_{t,t';\varphi} \{f_{t - \Tc
,t';\varphi} \}+
\lambda \mathcal{\hat{K}}_{t,t -\Tc;\varphi}\{1\} f_{t- \Tc,t';\varphi} \notag \\
& - \lambda \mathcal{\hat{K}}_{t,t';\varphi}\{ f_{t ,t'-
\Tc;\varphi} \}+ \lambda f_{t, t' - \Tc;\varphi}
\mathcal{\hat{K}}_{t' - \Tc,t';\varphi}\{1\} \, .
\end{align}
The integral kernel $\mathcal{\hat{K}}_{t,t';\varphi}\{F(\varphi)\}$
defined by Eq.~\eqref{Spectrum49}  is the generalization of the corresponding
differential operator derived in \oncite{VA03} for small angle
scattering.

We briefly discuss the term in the kinetic equation,
\req{Collision32full}, representing the electron-electron
interaction, ${\mathrm{St}}_{\rm ee}\{f\}$. As it was shown in
\oncite{DVAMP}, electric fields produce an isotropic non-equilibrium
contribution to the distribution function, which can only be
stabilized by the inelastic relaxation mechanisms. In weak electric
fields, the important inelastic relaxation is due to the
electron-electron scattering. To take it into account, we keep the
corresponding collision integral \req{Collision32}, which can be
written in the energy representation for a steady in time
distribution function  $ f (\vare)$: \be
\begin{split}
&{\rm St}_{\rm ee}\left\{ f (\vare) \right\} =
\int d\vare'\int d\Omega M(\Omega, \vare, \vare')\\
&\times \left[ \tilde f (\vare)f(\vare_+)\tilde f (\vare')
f(\vare'_-) -f(\vare)\tilde f (\vare_+)f(\vare') \tilde f (\vare'_-)
\right],
\end{split}
\label{Stin} \ee where $\tilde f(\vare) \equiv 1-f(\vare)$,
$\vare_{+}= \vare + \Omega$, $\vare_{-}'= \vare' - \Omega$ and
$M(\Omega,\,\vare,\,\vare')$ describes the dependence of the matrix
element of the screened Coulomb interaction on the transferred
energy $\Omega$ and the electron energies $\vare$ and $\vare'$. The
kernel $M(\Omega,\,\vare,\,\vare')$ has been discussed in details in
\oncite{DVAMP}. Below we use the linearized version of \req{Stin}:
\begin{equation}
{\rm St}_{\rm ee}\left\{ f (\vare) \right\} = -
\frac{f(\vare)-f_T(\vare)}{\tau_{\rm ee}}\, .
\end{equation}
Here, $\tau_{\rm ee}$ is the inelastic relaxation time due to the
electron-electron interaction, \be \frac{1}{\tau_{\rm ee}}\propto
\frac{T^2}{4\pi \vare_{\rm F}} \ln\frac{\kappa v_{\rm F}}{{\rm
max}\{T,\sqrt{\wc^{3}\tau_{\rm tr}}\}}\, ,\label{tauee=} \ee
which can be obtained as a projection of the linearized
electron-electron collision integral on the oscillating harmonic of
the distribution function, see \req{DF15iso} below.

\subsection{Bilinear response in microwave field}
\label{sec:Kin-Eq-Bilinear}

In this subsection we simplify the integral kernel
$\mathcal{\hat{K}}_{t,t'}$ in the limit of weak microwave power,
keeping only terms which are bilinear in the microwave electric
field, \emph{i.e.} linear in power $\Pmw$, introduced in \req{Pmw=}.
In the presence of microwave radiation $\mathcal{\hat{K}}_{t,t'}$
has the oscillatory dependence on the time variable $(t + t')/2$.
The distribution function in turn acquires non-stationary
corrections oscillating with the microwave frequency $\omega$. It
follows from the form of the kinetic equation,
\req{Collision32full}, that those corrections are small in the
parameter of the order of $ 1 / \omega \tau_{\rm tr} $ in systems
with considered here mixed disorder. We therefore neglect their
contribution to the distribution function and consider only the
stationary component of the distribution function. The latter  can
be found from the following equation
\begin{align} \label{Collision32}
\wc \partial_{\varphi } f_{t - t'}(\varphi) = \overline{\mathrm
{St}}_{\rm dis} \{ f \}_{t - t'} +\mathrm{St}_{\rm ee} \{ f \},
\end{align}
where the symbol $\overline{(\ldots)}$ stands for time averaging
over one period of the microwave oscillations and we included the
collision term due to the electron-electron interaction.
Due to the $2\pi/\omega$ periodicity of the collision integral,
\req{Collision36},   the time-average of the kernel
$\mathcal{\hat{K}}$, \req{Spectrum49}, is given by the integral over
one period of the microwave field.

We consider the response of a two-dimensional electron gas
to the in-plane electric field
\begin{align} \label{K41}
\vec{E}_{\mathrm{tot}} = \vec{E} + \vec{E}_{\mathrm{mw}} \, ,
\end{align}
represented as a superposition of a constant electric field $\vec{E}$
and a circularly polarized microwave field
\begin{align} \label{K43}
\vec{E}_{\mathrm{mw}} = E_{\omega} \mathrm{Re}\left[ \vec{e}^{\pm}
e^{ - i \omega t } \right] \, ,
\end{align}
where we introduced the complex polarization vector $\vec{e}^{\pm} =
(\vec{e}_x \pm i \vec{e}_y)/\sqrt{2} $  with the property
$\hat{\epsilon}\vec{e}^{\pm} = \pm i \vec{e}^{\pm}$. In the above
equations, the upper (lower) sign corresponds to the right (left) polarization
of the microwave field propagating in the magnetic field direction, see
Fig.~\ref{fig:shift}.

The displacement $\vec{\zeta}_{t_1,t_2}$, Eq.~\eqref{KinEq46b}, is
found by solving Eq.~\eqref{KinEq39} with the electric field
specified by Eqs.~\eqref{K41} and \eqref{K43}. The linearity of
Eq.~\eqref{KinEq39} allows us to represent its solution as the sum
of the displacements in constant and microwave fields,
\begin{align} \label{K61}
\vec{\zeta}_{t_1,t_2} = \vec{\zeta}^{\mathrm{dc}}_{t_1 - t_2} +
\vec{\zeta}^{\mathrm{ac},\pm}_{t_1,t_2},
\end{align}
where
\begin{align} \label{K68}
\vec{\zeta}^{\mathrm{dc}}_{t_1 - t_2} = \frac{ e (t_2 - t_1 )
\hat{\epsilon} \vec{E }}{ m \wc }\, ,
\end{align}
and
\begin{align} \label{K72}
\vec{\zeta}^{\mathrm{ac},\pm}_{t_1,t_2}= & \frac{ 2
\sqrt{\Pmw}}{\pF} \sin \frac{\omega(t_2 - t_1)}{2}
\mathrm{Im} \left[ \vec{e}_{\pm} e^{ - i \omega (t_1 + t_2)/2 }
\right]\, .
\end{align}
The dimensionless parameter $\Pmw$ has been
introduced in Eq.~\eqref{Pmw=}.
We make an expansion of the integral kernel
$\widehat{\mathcal{K}}_{t,t'}$ in Eq.~\eqref{Spectrum49} to the first order
in $\Pmw$ using the relation
\begin{align} \label{K81}
e^{ i \pF \left( \vec{n}_{\varphi} - \vec{n}_{\varphi'}
\right)\vec{\zeta}_{t_1,t_2} }
& \approx e^{ i \pF \left( \vec{n}_{\varphi} - \vec{n}_{\varphi'}
\right)
\vec{\zeta}^{\mathrm{dc}}_{t_1 - t_2} }
\notag \\
& \times \left[ 1 - p^2_F \overline{\left(\left( \vec{n}_{\varphi} -
\vec{n}_{\varphi'}
\right)\vec{\zeta}^{\mathrm{ac}}_{t_1,t_2}\right)^2} \right] \, .
\end{align}
Averaging \req{K81} with respect to time  results in
\begin{align} \label{K85}
\overline{\left[\left( \vec{n}_{\varphi} - \vec{n}_{\varphi'}
\right)\vec{\zeta}^{\mathrm{ac}}_{t_1,t_2}\right]^2} =
\frac{2 \Pmw }{ \pF^2 } \left(\vec{n}_{\varphi } -
\vec{n}_{\varphi'} \right)^2
\sin^2 \frac{ \omega ( t_1 - t_2 ) }{2} \, .
\end{align}
And finally the collision kernel to the second order in the
microwave field takes the form
\begin{equation}
\begin{split}
\label{K89}
\mathcal{\hat{K}}_{t,t';\varphi}\{F(\varphi)\}& = \int\frac{d\varphi'}{2\pi}
%& \int_0^{2\pi} \frac{d\varphi'}{2\pi}
%
e^{ iW_{\varphi\varphi'}\cdot(t - t') }F(\varphi') \\
\times & \left[ \frac{ 1 }{\tau_{\varphi' - \varphi} }\!-\!\Pmw
\frac{ 1 - \cos\omega(t -
t')}{\bar{\tau}_{\varphi' - \varphi}}%
\right]\, ,
%e^{ (\varphi' - \varphi)\partial_{\varphi} } \, ,
\end{split}
\end{equation}
where the rate $1/ \bar{\tau}$ has been introduced in
Eq.~\eqref{qualit37}
and the quantity
\begin{align} \label{K93}
W_{\varphi \varphi'} = e E \Rc (\sin \varphi - \sin \varphi' )\,
\end{align}
is the work done by the dc electric field as the result of
scattering off an impurity,\cite{VAG} $W_{\varphi \varphi'} = e
\vec{E} \Delta \vec{R}_{\varphi \rightarrow \varphi'}$ with the
shift of the cyclotron orbit $\Delta \vec{R}_{\varphi \rightarrow
\varphi'}$ given by Eq.~\eqref{qualit13}.

The collision integral due to scattering off disorder is obtained
from \req{K89} by performing the Fourier transformation in time
variable $t-t'$. We represent it as the sum of two terms describing
two separate scattering mechanisms
\begin{align} \label{K113}
\overline{\mathrm{St}}_{\rm dis} = \mathrm{St}_{\rm dc} + \mathrm{St}_{\rm mw}
\, .
\end{align}
In \eqref{K113} the first term corresponds to the scattering off the
impurities in the absence of the microwave radiation\cite{VAG}
\begin{align} \label{K123}
\mathrm{St}_{\rm dc}f = &
\int \frac{ d \varphi'}{ 2 \pi }\frac{ \nu(\vare +
W_{\varphi\varphi'})}{ \nu_0 }
\frac{ f( \vare + W_{\varphi\varphi'} , \varphi' ) -
f( \vare , \varphi ) }{ \tau_{\varphi' - \varphi} }\, .
\end{align}
The second term in \req{K113}
\begin{widetext}
\begin{align} \label{K127}
\mathrm{St}_{\rm mw}f = &
- \frac{\Pmw}{ 2 \nu_0} \sum_{\pm}
\int \frac{ d \varphi'}{ 2 \pi } \frac{ 1 }{ \bar{\tau}_{\varphi' -
\varphi} } \left\{
\nu(\vare\! +\! W_{\varphi\varphi'})\left[ f(\vare +
W_{\varphi\varphi'}, \varphi') - f(\vare, \varphi) \right]
\notag
\right. \\
& \left. - \nu(\vare\! +\! W_{\varphi\varphi'} \pm \omega) \left[
f(\vare + W_{\varphi\varphi'} \pm \omega, \varphi') - f(\vare,
\varphi) \right]
\right\}
\end{align}
\end{widetext}
describes the scattering processes of electrons off impurities with
participation of a microwave radiation quantum. For the analysis
limited to the first order in the microwave power, all multi-photon
processes are neglected. More specifically, the
first term in \req{K127} represents the impurity scattering with
one photon emitted (absorbed) virtually and can be thought of as the
renormalization of the impurity potential by microwave radiation.
This term is taken into account in Sec.~\ref{sec:Qualit-StrongDC} as
a linear in $\Pmw$ contribution to $j_{\rm el}$ in \req{qualit15}.
The second term in \req{K127} describes the real processes of emission
(absorption) of one microwave quantum accompanying the impurity
scattering. It is this term which produces the oscillatory
$\omega/\wc$ dependence of the magneto-resistance and corresponds to
$j_{\rm in}$ in \req{qualit15}.

We notice that the obtained collision integral vanishes in the clean
system for any frequency $\omega$ away from the cyclotron resonance.
In the clean limit, the conductivity tensor can be found by applying
the Kohn's theorem\cite{Kohn} argumentation. The conductivity tensor
of an interacting system, which is  galilean invariant,
is identical to that of the non-interacting system.\cite{ChiralRes}
It follows then that away from the cyclotron resonance, $\omega \neq
\omega_c$, the electric field appears in the collision integral only
in a combination with the disorder scattering rate $1/\tau_{\theta}$.

\section{Magneto-oscillations in the presence of ac and dc
excitations} \label{sec:ACDC}

In this section we calculate and analyze the dissipative current
\begin{align} \label{K132}
j= 2 e v_F \int \frac{ d \varphi }{ 2 \pi } \cos \varphi \int
\nu(\vare) f(\vare, \varphi) d \vare\, ,
\end{align}
where the distribution function is determined as a solution of the
kinetic equation \req{Collision32} with the collision integral
given by \reqs{K113}, \eqref{K123} and \eqref{K127}.

\subsection{Solution of the kinetic equation}
\label{sec:ACDC-Solv-Kin-Eq}
We look for the solution of the kinetic equation,
Eq.~\eqref{Collision32}, in the form
\begin{align} \label{DF13}
f(\vare, \varphi) = f_T(\vare) + \delta f_{\mathrm{cl}}(\vare, \varphi) +
\delta f_0(\vare) + \delta f_1(\vare, \varphi)\, ,
\end{align}
where the first term is the equilibrium Fermi-Dirac distribution
function, \req{ft=}. The second term is the classical solution corresponding to
the constant density of states
\begin{align} \label{DF14}
\delta f_{\mathrm{cl}}(\vare, \varphi) = - \partial_{\vare} f_T \frac{ e E
R_c }{ \wc \tautr } \cos \varphi\,,
\end{align}
leading to the Drude result for the longitudinal conductivity at
large Hall angle, \req{qualit45}.

The third and fourth terms in Eq. \eqref{DF13} are the zeroth and
first angular harmonics of the correction to the distribution
function resulting from the quantum oscillatory component of the
density of states in the collision integral \eqref{K113}. For $\wc
\tau_{\rm tr}\gg 1 $, we keep only the isotropic component and the
first angular harmonic of the distribution function:
\begin{subequations} \label{DF15}
\begin{eqnarray}
\delta f_0 (\vare) & = & \lambda \partial_{\vare}f_T I
\sin \frac{ 2 \pi \vare}{ \wc } \label{DF15iso} \\
\delta f_1 (\vare,\varphi) & = & \lambda
\partial_{\vare}f_T \left[
A_1 \cos \frac{ 2 \pi \vare}{ \wc } + \lambda A_2 \right] \cos
\varphi\, . \label{DF15ani}
\end{eqnarray}
\end{subequations}
The coefficients $I$, $A_1$ and $A_2$ are fixed by the kinetic
equation~\req{Collision32}. The calculation of these coefficients is
outlined in the Appendix.

The amplitude $I$ of the isotropic part of the distribution function
is
\begin{equation}
\begin{split}
\label{DF31} I = & -\frac{\wc}{\pi}\frac{1}{ \tau_{\rm ee}^{-1} +
\tau_0^{-1 } - \gamma(\edc) + 2\Pmw \bar{\gamma}(\edc) \sin^2\pi\eac
}
\\
\times & \Big[
 \edc \gamma'(\edc) - 2 \pi\eac \sin 2\pi\eac \Pmw \bar{\gamma}(\edc)
\\
& - 2\edc \sin^2 \pi\eac   \Pmw \bar{\gamma}'(\edc) \Big]\, .
\end{split}
\end{equation}
Here, functions $\gamma(\edc)$ and $\bar \gamma(\edc)$
are defined as
\begin{subequations}
\label{DF30}
\begin{eqnarray} \label{DF30nobar}
\gamma(\edc) =  \sum_n \frac{ J_n^2(\pi\edc) }{ \tau_n }\,
\end{eqnarray}
and
\begin{align}
\bar \gamma(\edc) = & \sum_n J^2_n(\pi\edc)
\left[\frac{ 1 }{ \tau_n }-\frac{ 1 }{ 2\tau_{n+1} }-\frac{ 1 }{ 2\tau_{n-1}
}\right]
\label{DF30bar}
\end{align}
\end{subequations}
in terms of the Bessel functions $J_n(\pi \edc)$ of the
order $n$ and angular harmonics $1/\tau_n$ of scattering rate off disorder, \req{taundef=}.
Term $\tau_0^{-1}$ in the denominator of \req{DF31} is the
$n=0$ harmonic of scattering rate off disorder and coincides with
the quantum scattering rate off disorder.
Coefficient $I$ contains the inelastic relaxation time $\tau_{\rm ee}$, \req{tauee=}.
Equation~\rref{DF31} was obtained in \oncite{VAG}
for arbitrary value of $\edc$ and $\Pmw=0$ and in \oncite{DVAMP} for
$\edc\ll 1$ and $\Pmw\ll 1$ and has been shown to be important for
the description of MIRO and HIRO at small voltages.
Equation \eqref{DF31} determines the
behavior of the isotropic non-equilibrium component of the
distribution function in constant electric field of an arbitrary strength.

The amplitude $A_1$  is given by
\begin{equation}
\begin{split}
\label{DF35} A_1 =& -\frac{ I }{\pi  \omega_c } \left( \gamma'(\edc)
- 2 \Pmw
\bar{\gamma}' (\edc)  \sin^2\pi \eac  \right) \\
& + \frac{ 1}{ \pi^2}
\Big( \edc \gamma''(\edc)  - 2 \pi \eac \sin 2 \pi \eac
\Pmw  \bar{\gamma}'(\edc)   \\
&
- 2 \edc \sin^2\pi \eac \Pmw   \bar{\gamma}''(\edc)
\Big)
 \, .
\end{split}
\end{equation}
And finally we obtain the amplitude $A_2$ in the following form
\begin{align} \label{DF41}
A_{2} = \frac{ I }{\pi \wc }\left( \gamma'(\edc) - 2 \Pmw\bar{
\gamma }'(\edc)\sin^2\pi \eac \right)\, .
\end{align}
Equations \eqref{DF31}, \eqref{DF35} and \eqref{DF41} determine the
first two angular harmonics of the distribution function through
Eq.~\eqref{DF15}. This allows us to compute the current density as
discussed in the next subsection.

\subsection{Non-linear current}
\label{sec:ACDC-NL-Curr}

In this section we calculate  the dependence of the current  to the first order in power $\Pmw$
on parameters $ \edc $ and $\eac$, characterizing the strength of the dc
and ac excitations respectively. We
substitute  the distribution function \req{DF13} with factors $I$
and  $A_{1,2}$ given by \reqs{DF31}, \rref{DF35} and \rref{DF41} to
the expression for the dissipative current, \req{K132}:
\begin{align}
\label{current2=}
j\left(\edc, \eac \right) = \sigma_{\mathrm D} E + \delta j \, ,
\end{align}
where the Drude conductivity $\sigma_{\mathrm D} $
is given by \req{qualit45} and
the leading correction in $\lambda^2$ to the current has the form
\begin{equation} \label{MagnetoR9}
\delta j= \lambda^2 \left( A_1 - A_2 \right) e v_F \nu_0 \, .
\end{equation}

Substituting $A_{1,2}$ from \reqs{DF35} and
\rref{DF41} to \req{MagnetoR9}, we represent the correction to the current  in terms
of the dimensionless function $F \left( \edc,\eac \right)$:
\be
\label{deltaj=}
\delta j= 2\sigma_{\rm D} E \lambda^2 F(\edc,\eac).
\ee
The function $F(\edc,\eac)$ in Eq.~\eqref{deltaj=} describes the
leading correction to the classical value of the current due to
oscillations of the electron density of states. This function can be
represented as a combination of the ``displacement'' term $\fnd
\left(  \edc,\eac \right)$ and the ``inelastic'' term $\fni \left(  \edc,\eac \right)$, arising due to the non-equilibrium
isotropic component of the distribution function:
\be
\label{MagnetoR14}
F \left(
\edc,\eac \right) = \fnd \left(  \edc,\eac \right) + \fni
\left(  \edc,\eac \right)\, .
\ee
The ``displacement'' contribution $\fnd$ originates from the
modification of the scattering rates in crossed electric and magnetic fields,
while neglecting the effect of electric fields on the isotropic component of the
electron distribution function.
We have
\begin{equation}
\bes
 \label{MagnetoR16}
\frac{\fnd}{ \tautr }=& -\frac{\gamma''(\edc)}{\pi^2}
\\
& + \frac{ 2
\Pmw}{ \pi^2 }
\left( \frac{ \pi \eac \bar{\gamma}'  \sin 2 \pi \eac}{ \edc }    +
 \bar{ \gamma}'' \sin^2 \pi \eac
 \right) \,
\end{split}
\end{equation}
with $\gamma=\gamma(\edc)$ and
$\bar\gamma=\bar\gamma(\edc)$ defined by  \reqs{DF30}.

The remaining ``inelastic'' contribution is proportional to
the amplitude $I$ of the isotropic and energy-dependent
non-equilibrium component of the distribution function $\delta f_0(\vare)$. This
contribution is sensitive to the energy relaxation rate $1/\tauin$ at
sufficiently small values of $\edc$:
\begin{equation}
\begin{split}
 \label{MagnetoR21}
\frac{\fni}{ \tautr }  = &
- \frac{2}{\pi^2 \edc}
\frac{ \gamma'  - 2 \bar{ \gamma }'  \Pmw  \sin^2 \pi \eac
    }{
     \tauin^{-1}\! + \!\tau_{0}^{-1}\! - \!\gamma  \! +\!
    2 \bar{ \gamma }  \Pmw \sin^2 \pi \eac
    } \\
 \times &
\Big( \edc \gamma'   -   2   \pi \eac  \bar{ \gamma}  \Pmw  \sin 2\pi \eac
%\\
%&
-2\edc \bar{\gamma}'  \Pmw   \sin^2 \pi \eac    \Big)
\, .
\end{split}
\end{equation}
Equations \eqref{deltaj=}, \eqref{MagnetoR14}, \eqref{MagnetoR16}
and \eqref{MagnetoR21} give the explicit expression for the current
in response to the applied dc electric field with strength
$E=\edc\wc/(2|e|\Rc)$ in weak microwave fields $\Pmw\ll 1$.

First, we analyze the properties of functions $\fnd(\edc,\eac)$ and
$\fni(\edc,\eac)$ in a weak dc electric field, $\edc\ll 1$. In
this case
\begin{subequations}
\label{gammalowE}
\begin{eqnarray}
\gamma(\edc) & = &
\frac{1}{\tau_0}-\frac{\pi^2\edc^2}{2\tautr}+\frac{\pi^4\edc^4}{32}\frac{1}{\tau_*}\,,
\label{gammalowEa}
\\
\bar \gamma(\edc) & = &
\frac{1}{\tautr}-\frac{1}{4}\pi^2\edc^2\frac{1}{\tau_*}\,,
\label{gammalowEb}
\end{eqnarray}
\end{subequations}
where
$1/\tau_0\equiv 1/\tau_{n=0}$ is the quantum scattering rate off
disorder,
\be
\frac{1}{\tautr}=\frac{1}{\tau_0}-\frac{1}{\tau_1}
\ee
is the transport
scattering rate written in terms of harmonics of scattering rate, see \reqs{qualit45} and
\rref{taundef=}, and
\be
\label{taustar}
\frac{1}{\tau_*}=\frac{3}{\tau_0}-\frac{4}{\tau_1}+\frac{1}{\tau_2}.
\ee
Substituting \reqs{gammalowE} to \reqs{MagnetoR16} and
\rref{MagnetoR21}, we obtain
\begin{equation}
\label{FdlowE=}
\fnd =  1-\frac{\tautr}{\tau_*}
\left[
    \frac{3}{8}\pi^2\edc^2+\Pmw
    \left(
        \pi\eac\sin 2\pi\eac+\sin^2\pi\eac
    \right)
\right]\,
\end{equation}
and
\begin{equation}
\begin{split}
\label{FnelowE=}
& \fni=-
2\frac{1-(\tautr/\tau_*)\Pmw\sin^2\pi\eac}{\tautr/\tauin+\pi^2\edc^2/2+2\Pmw\sin^2\pi\eac}\\
&\times
\left[
    \pi^2\edc^2\left(1-(\tautr/\tau_*)\Pmw\sin^2\pi\eac \right)
    +2\pi\Pmw \eac\sin 2\pi\eac
\right].
\end{split}
\end{equation}

In smooth disorder, $\tautr\ll\tau_*$. In this case, \req{FnelowE=}
coincides with the contribution to the electric current,
considered in \oncite{DVAMP}, which we already presented
in \req{qualit66}. The non-linear dependence on the
applied electric fields occur at $\edc\lesssim
\sqrt{\tautr/\tauin}$.

In strong dc electric fields functions $\gamma(\edc)$ and
$\bar\gamma(\edc)$ can be evaluated for $\edc\gg 1$
using the asymptotes of the
Bessel functions. As a result,  the  we find the ``displacement'' contribution
$\fnd(\edc,\eac)$ in the form:
\be
\begin{split}\label{FdgtrE=}
\frac{\fnd}{\tautr} = &\frac{4}{\pi^2}\frac{1}{\edc\tau_\pi}
    \sin 2 \pi\edc(1-2\Pmw) \\
&+
    \frac{4}{\pi^2} \frac{\Pmw }{\edc\tau_\pi}\sum_\pm
    \frac{\edc\pm\eac}{\edc}\sin2\pi(\edc\pm\eac)
\end{split}
\ee
with $1/\tau_\pi=\sum_{n}\cos\pi n/\tau_n$ being the back
scattering rate off disorder. The function $\fnd(\edc,\eac)$
has an oscillatory dependence on the parameter $\edc$. We presented
the corresponding expression for the non-linear contribution to the
current in Sec.~\ref{sec:Qualitative}, \reqs{qualit51} and
\rref{qualit56}, where the same result was obtained from
semi-qualitative arguments and the stationary phase approximation.

For the ``inelastic'' contribution $\fni$ in the limit
$\edc\gg 1$ we obtain the following estimate
\be
\begin{split}\label{FnegrtE=}
\frac{\fni}{\tautr}& = - \frac{8}{\pi^4}\frac{1}{\edc\tau_\pi}
\frac{\tau_0}{\tau_\pi\edc}
\cos^2 2\pi\edc
\\
&\times
\left[1-2\Pmw
    \left(
        4\sin^2\pi\eac+\frac{\eac}{\edc}\tan 2\pi\edc\sin2\pi\eac
    \right)
\right].
\end{split}
\ee
We notice that this contribution $\fni(\edc,\eac)$
is smaller than $\fnd(\edc,\eac)$ by  factor $(1/\edc)$
in strong electric fields $\edc\gg 1$. For a system with smooth
disorder $\fni(\edc,\eac)$ contains also an additional small factor
$\tau_0/\tau_\pi\ll 1$. This smallness
allows one to neglect the ``inelastic'' contribution to the
current in strong electric fields. The first line of
\req{FnegrtE=} coincides\cite{atypo} with the asymptote found in
\oncite{VAG}.
The second line of \req{FnegrtE=} describes the effect
of microwave field in the bilinear response.

\subsection{Model of disorder}
\label{sec:ACDC-Model}
The purpose of this section is to analyze our results for a
particular model of disorder. The disorder in GaAs/AlGaAs
heterostructures is mainly due to remote charged donors. The
potential created by these donors is smooth on the electron
wave-length scale. Therefore, such potential is characterized by an
exponentially suppressed backscattering amplitude. As we saw in
Sec.~\ref{sec:Qualit-StrongDC} the latter is crucial for the onset
of non-linear magneto-oscillations at higher current densities. On
the other hand the relatively weak in-plane disorder creates the
scattering potential of the short range giving rise to a finite
backscattering amplitude.

An adequate model should describe a mixed disorder that
includes impurities with narrow and wide angle
scattering.\cite{mirlindisorder}
Specifically, we use the expression for the angular harmonics of the
scattering rate employed in \oncite{VAG}
\begin{align} \label{DisorderModel13}
\frac{ 1 }{ \tau_n } = \frac{ \delta_{n,0} }{ \taus }
+\frac{ 1 }{ \taul }\frac{ 1 }{ 1 + \chi n^2}
\, .
\end{align}
For this model  we have the following expressions for the quantum
and transport scattering rates \be \label{tautrmodel=}
\frac1\tauq=\frac1{\tau_0}=\frac{1}{\taus}+\frac{1}{\taul},\quad
\frac1\tautr=\frac1\taus+\frac{1}{\taul}\frac{\chi}{1+\chi}. \ee We
perform calculations under the assumptions: $\wc\tauq\lesssim 1$ and
$\wc\tautr\gg 1$, which can be met for disorder described by
\req{DisorderModel13}, provided that $\chi\ll 1$ and
$\taus\gg\taul$. The last inequality means that the long-range
disorder is stronger than the short-range disorder. The scattering
time $\tau_*$ is given by \be \label{taustarmodel=}
\frac{1}{\tau_*}=\frac{3}{\taus}+ \frac{12\chi^2}{\taul}. \ee
Equations~\rref{tautrmodel=} and \rref{taustarmodel=} determine the
ratio $\tautr/\tau_*$, which appears in \reqs{FdlowE=} and
\rref{FnelowE=} for the non-linear contributions to the current at
weak electric fields.

For very small values of $\edc$ and $\Pmw$, we have the following
estimate \be \fni=-\frac{2\tau_{\rm ee}}{\tautr}\left(
\pi^2\edc^2+2\pi\Pmw\eac\sin 2\pi\eac \right). \ee Comparison of
this expression with \req{FdlowE=} shows that the linear in $\Pmw$
and $\edc^2$ contributions to the non-linear current are dominated
by the ``inelastic'' mechanism if $\tauin/\tautr\gg\tautr/\tau_*$.
If only long range disorder is present in the system, we obtain for
the ratio $\tautr/\tau_*=12\chi \ll 1$ and the ``inelastic''
mechanism is indeed dominant for sufficiently low temperatures,
while $\tauin\gtrsim\tautr\chi\sim \tau_0$. At higher temperatures
the interaction effects suppress the oscillations of the density of
states and decrease the overall amplitude of non-linear current. In
the case of mixed disorder, parameter $\tautr/\tau_*$ could be of
the order of unity and the ``inelastic'' and ``displacement''
mechanisms become comparable at lower temperatures, when
$\tauin\sim\tautr$, but still $\tauin\gg\tau_0$ and the oscillations
of the density of states are not smeared by the interaction effects.
As a consequence, there exists a regime where the linear
photo-resistivity may have significantly weak dependence on electron
temperature.\cite{zudov-pc}

We use the model for disorder described by \req{DisorderModel13} to
evaluate the back scattering rate, which characterizes non-linear
contributions to the current in strong electric fields $\edc\gg 1$.
Performing summation over index $n$ for $\theta=\pi$ in
\req{taundef=}, we find \be\label{taupi=}
\frac{1}{\tau_\pi}=\frac{1}{\taus}+
\frac{1}{\taul}\frac{2\pi}{\sqrt{\chi}}\exp\left(-\frac{\pi}{\sqrt{\chi}}\right)\,
. \ee
This rate controls the magneto-oscillations for $\edc \gg 1$,
\reqs{FdgtrE=}  and \rref{FnegrtE=}. For small values of $\chi$, the
second term can be disregarded.

\begin{figure}[h]
\includegraphics[width=0.85\columnwidth]{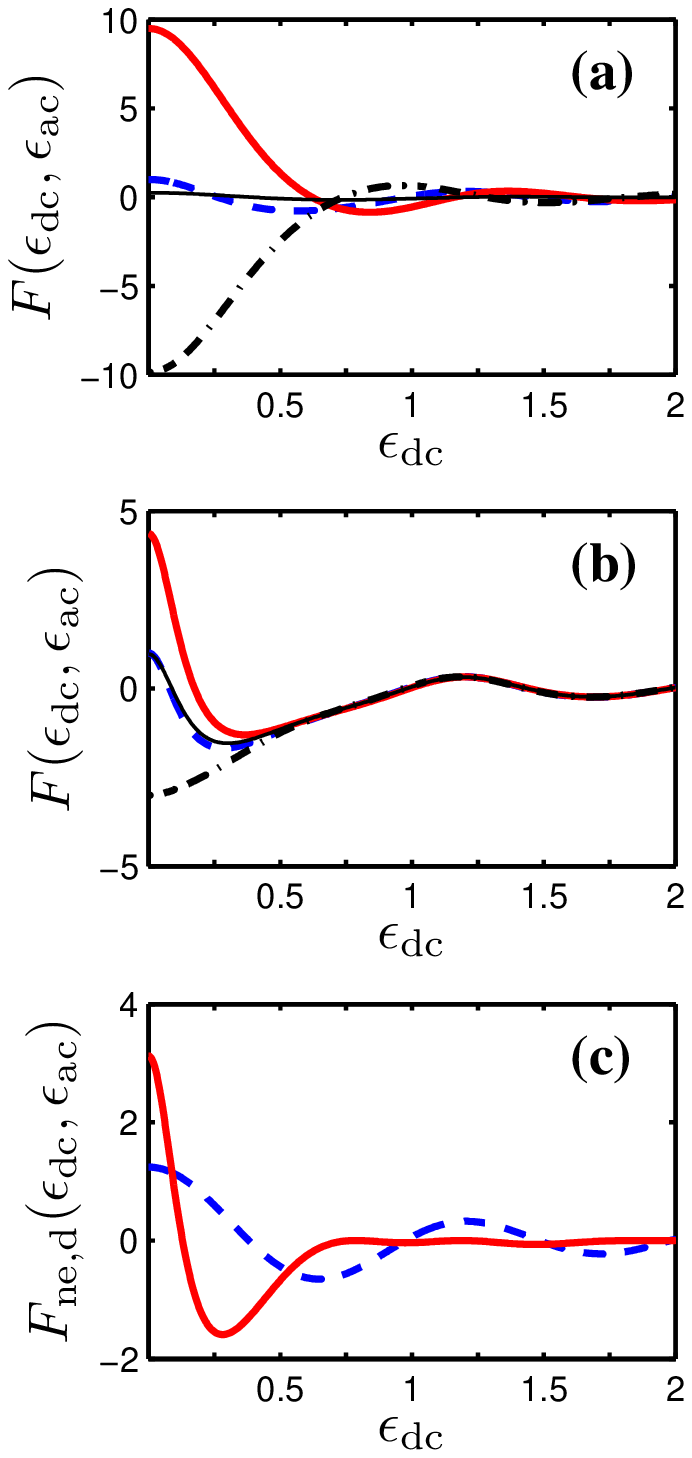}
%\vspace{-0.5cm}
\caption{(Color online) The function $F(\edc,\eac)$ as defined by
Eqs.~\eqref{MagnetoR14}, \eqref{MagnetoR16}, and \eqref{MagnetoR21}
with  $\gamma(\edc)$ and $\bar{\gamma}(\edc)$ given by
Eqs.~\eqref{gammainmodel=} for the model of
disorder introduced in Sec.~\ref{sec:ACDC-Model}.
(a) $\tau_{\mathrm{sm}} = 1$, $\tau_{\mathrm{sh}} = 1$,
$\tau_{\mathrm{ee}} = 0.5$, $\chi = 0.0001$ and $\Pmw = 0.25$ ;
(b) $\tau_{\mathrm{sm}} = 0.1$, $\tau_{\mathrm{sh}} = 1$,
$\tau_{\mathrm{ee}} = 10$, $\chi = 0.001$ and $\Pmw = 0.01$.
In both cases the thick solid (red) line, the dashed (blue) line,
the dashed-dotted (black) line and the thin solid (black) line are
used for $\eac = 2.75 $, $\eac = 3 $, $\eac = 3.25 $, and $\eac =
3.5 $, respectively.
The panel (c) shows the inelastic contribution, $F_{\mathrm{ne}}$,
solid (red) line, and displacement contribution $F_{\mathrm{d}}$,
dashed (blue) line for the set of parameters of panel (b) at
$\eac = 2.75$.
 \label{fig:F}}
\end{figure}

To analyze the case of arbitrary $\edc$, we evaluate functions
$\gamma(\edc)$ and $\bar\gamma(\edc)$,  \reqs{DF30}, in the limit
$\chi\ll 1$. We perform summation over index $n$ in \req{DF30} using
the Poisson formula and neglect exponentially small corrections,
similar to the second term in \req{taupi=}. The result
reads\cite{VAG}
\begin{subequations}\label{gammainmodel=}
\begin{align} \label{DisorderModel17}
\gamma(\edc) =  &\frac{ J_0^2(\pi\zeta) }{ \taus }+
\frac{ 1 }{ \taul }\frac{ 1 }{ \sqrt{1 + \chi \pi^2\edc^2} }\,,
%%\end{align}
%%%
%%and
%%%
%%\begin{equation}
%%\begin{split}
\\
\label{DisorderModel20}
\bar{\gamma}(\edc) = &
\frac{ 1 }{ \taus } \left( J_0^2(\pi\edc) - J_1^2(\pi\edc) \right)
\\
& +
\frac{ \chi }{\taul }
\frac{ 1 - \chi \pi^2\edc^2 / 2}{ (1 + \chi \pi^2\edc^2)^{5/2} }\,
\nonumber
.
\end{align}
%\end{equation}
\end{subequations}
Equations \rref{gammainmodel=} are convenient to calculate the
non-linear contribution to the current at arbitrary value of $\edc$.
The resulting function $F(\edc,\eac)$, \req{MagnetoR14}, is plotted
in Figs.~\ref{fig:F}(a,b). We also compare contributions $\fnd(\edc,\eac)$ and $\fni(\edc,\eac)$ to the net non-linear
current. Both functions are shown for the same set of parameters in
Fig.~\ref{fig:F}(c). The ``inelastic'' contribution $\fni(\edc,\eac)$ has a large variation at small $\edc$, and vanishes
at larger values of $\edc$. On the other hand, the displacement
component $\fnd(\edc,\eac)$ is an oscillatory function of the
parameter $\edc$ with only weakly decaying amplitude.

For $\edc\gtrsim 1$, when the ``inelastic'' contribution
becomes small, the function $F(\edc,\eac)$ coincides with the
``displacement'' contribution $\fnd(\edc,\eac)$. The latter
is a sum of two terms. One term contains $1/\taul$, this term
varies smoothly as a function of $\edc$ on the scale $\edc\sim
1/\sqrt{\chi}$. Another term contains $1/\taus$ and oscillates as
a function of $\edc$ with period equal to unity. The corresponding
oscillating term can be written as
\begin{equation}
\begin{split} \label{DisorderModel23}
\frac{F_{\mathrm{d,osc}}}{\tautr} = &
-\frac{ [J_0^2(\pi\edc)]'' }{\pi^2 \taus }
\\
& + \frac{ 2\Pmw }{ \pi^2 \taus }
\Big[(J_0^2(\pi\edc) - J_1^2(\pi\edc))''\sin^2\pi\eac
\\ &
+ \frac{\pi \eac}{ \edc }(J_0^2(\pi\edc) - J_1^2(\pi\edc))' \sin  2 \pi \eac \Big] \, .
\end{split}
\end{equation}
This equation indicates clearly the vital importance of the
short-range disorder, characterized by the finite backscattering
rate $1/\taus$ for the onset of the magneto-oscillations in the
non-linear transport regime.

\subsection{Differential magneto-resistance}
\begin{figure}[h]
\includegraphics[width=0.85\columnwidth]{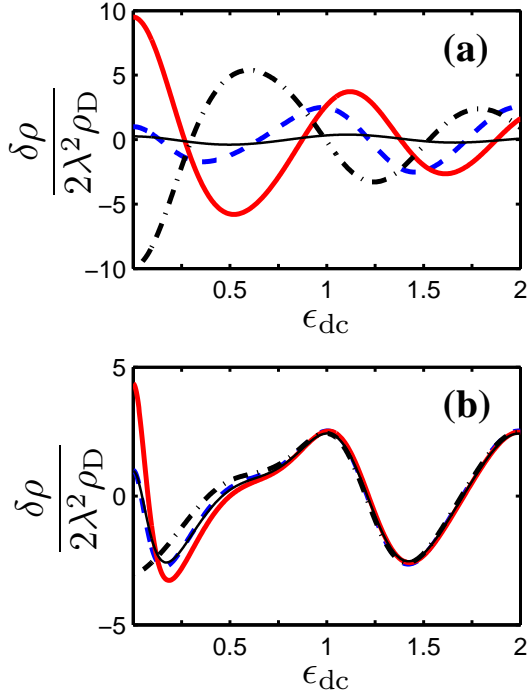}
%\vspace{-0.5cm}
\caption{(Color online) The upper and lower panels show
the differential resistivity as a function of $\edc$
for values of parameters used in Figs.~\ref{fig:F}(a) and (b),
respectively. The differential resistance is
obtained by substitution of $F(\edc,\eac)$, shown in Figs.~\ref{fig:F}(a) and (b),
to Eq.~\eqref{MagnetoR41}.
\label{fig:rho}}
\end{figure}

Now we apply \req{current2=} for the electric current response to
the applied dc electric field to describe the longitudinal
differential magneto-resistance
\begin{align} \label{MagnetoR24}
\rho(j) = \partial E_{\parallel}/ \partial j\,.
\end{align}
In Eq.~\eqref{MagnetoR24} we have chosen $\vec{e}_x$ to be the current direction,
so that $E_{\parallel}$ is the electric field component parallel to
the current. In the limit $\wc \gg 1 / \tautr $ we write
\begin{equation} \label{MagnetoR28}
E_{\parallel} =\rho_{\rm D} j \, , \quad \rho_{\rm
D}=\frac{1}{e^2\nu_0 \vF^2\tautr}\, ,
\end{equation}
where the electric current density $j$ is given by \req{current2=}
\begin{equation}
\label{MagnetoR37} E_{\parallel} = \rho_{\rm D} j \left[ 1 + 2
\lambda^2 F( \edc^j, \eac)  \right]
\end{equation}
calculated with a total electric field $E\simeq \rho_{\rm H}j$,
where $\rho_{\rm H}=\wc/(e^2\vF^2\nu_0)$ is the Hall resistance and
\begin{equation}
\label{edcj}
\edc^j  = \frac{2|e| (\rho_{\rm H}j) \Rc}{\wc}=\frac{4\pi j}{ep_{\rm
F}\wc}.
\end{equation}
We finally obtain for the oscillatory part of the differential
resistance, $\delta \rho = \rho - \rho_{\rm D}$ the following expression
\begin{align} \label{MagnetoR41}
\frac{ \delta \rho(j) }{ \rho_{\rm D} } = & 2 \lambda^2 \frac{ d }{
d \edc^j }\left[ \edc^j F( \edc^j, \eac ) \right] \,
\end{align}
with function $F( \edc^j, \eac )$  given by \reqs{MagnetoR14},
\rref{MagnetoR16} and \rref{MagnetoR21}.
Figure~\ref{fig:rho} shows the differential resistance found for
the specific disorder model considered in
Sec.~\ref{sec:ACDC-Model}.

We analyze our results in the limiting cases of small and large dc
currents. We start with the discussion of the differential
resistance $\delta \rho$ at small direct current, $\edc^j\lesssim
1$, when in samples with smooth disorder
the non-linear behavior originates from $\fni(
\edc^j, \eac )$ contribution. Keeping only this contribution in
\req{MagnetoR41}, at $\edc^j=0$ we have
\begin{align} \label{MagnetoR44}
\frac{ \delta \rho }{ \rho_D } = & -2 \lambda^2 \frac{
 2 \pi \eac \Pmw \sin 2 \pi \eac }
{\tautr/ \tauin + 2 \Pmw  \sin^2 \pi \eac }\,.
\end{align}
As $\edc^j$ increases, the differential resistance decreases from the
above value on the scale $\edc^j\simeq\sqrt{\tautr/\tauin}$ and
exhibits a non-monotonic behavior at $\edc^j\gtrsim 1$.
However for large values of $\edc^j$, the
``inelastic'' contribution to the differential resistance
vanishes fast, as discussed in Sec.~\ref{sec:ACDC-NL-Curr}.

We now analyze the differential resistance at large values of direct current,
$\edc^j\gg  1$. In this limit the ``displacement'' contribution  dominates over
the ``inelastic'' contribution. Therefore, we keep only $\fnd(\edc^j,\eac)$ and find
\begin{equation}
\begin{split}
\label{MagnetoR49}
 &\frac{ \delta \rho }{ \rho_D }=  \frac{ ( 4 \lambda )^2
\tautr }{\pi \tau_\pi  }\bigg[
( 1 - 2 \Pmw ) \cos 2 \pi \edc
\\
&
+ 2 \Pmw \left(\cos 2 \pi \edc \cos 2 \pi \eac
%
%%%%%\right. \notag \\
%%%%%& \left.
- \frac{ \eac }{ \edc } \sin 2 \pi \edc \sin 2 \pi \eac
\right) \bigg] \, .
\end{split}
\end{equation}
The first term in square brackets coincides with the result of
\oncite{VAG} at $\Pmw=0$. At finite $\Pmw$ the factor $( 1 - 2 \Pmw )$
represents the impurity potential renormalization due to the virtual absorption
and emission of photons. These radiative corrections tend to
suppress the dark resistivity and
can be interpreted as motional narrowing or averaging out the electrostatic potential
of impurities in the presence of an oscillating electric field.
The other  two terms in \req{MagnetoR49}
describe the effect of the combined scattering off disorder in mixed constant
and oscillating electric fields when the real
microwave photons are absorbed or emitted.

\begin{figure}[h]
\includegraphics[width=1.0\columnwidth]{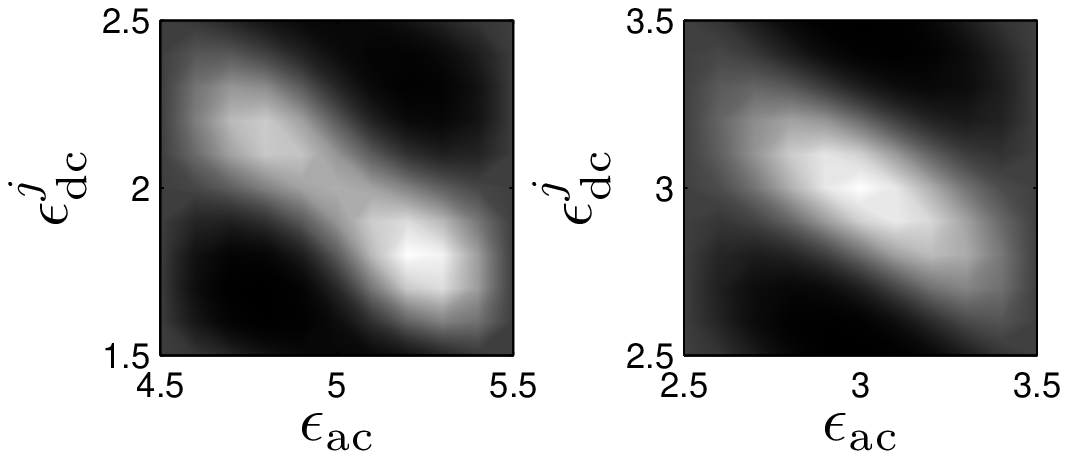}
\caption{(Color online) The differential resistance obtained from
Eq.~\eqref{MagnetoR49} for $\Pmw=0.25$ presented on a grey-scale
plot. The bright areas correspond to a higher value of the
resistance. (left) Away from the main diagonal in the $
(\eac,\edc^j) $ plane the differential resistance has minima and
maxima described by the conditions~\eqref{MagnetoR51}. (right) Close
to the main diagonal $ \eac \simeq \edc^j $, the differential
resistance becomes a function of the sum $ \eac + \edc^j $.
 \label{fig:color}}
\end{figure}

We briefly discuss the properties of \req{MagnetoR49} as a function
of two parameters $\edc^j$ and $\eac$. It is plotted in
Fig.~\ref{fig:color} as a grey-scale contour plot. This function
exhibits a series of maxima and minima in $(\eac, \edc^j)$ plane.
In general, for $\eac > \edc^j$ the local maxima and minima of the
function defined by \req{MagnetoR49} are located at
%
%\begin{subequations}
\begin{align}\label{MagnetoR51}
( \eac ,\edc^j )^{\mathrm{max}} & = (m \pm 1/4 , n \mp 1/4)
%\end{align}
\notag \\
%
%\begin{align}
( \eac,\edc^j )^{\mathrm{min}} & = (m \pm 1/4 , n \pm 1/4) \, .
\end{align}

We further notice that in the region of parameters $\left| \eac -
\edc^j \right| \lesssim \left| \eac + \edc^j \right|$, namely not to
far from the main diagonal in the two-dimensional plain $
(\eac,\edc) $ the result \eqref{MagnetoR49} reduces to
\begin{equation}
\begin{split}
\label{MagnetoR52}
\frac{ \delta \rho }{ \rho_D } \approx & \frac{ ( 4 \lambda )^2
\tautr }{\pi \tau_\pi }
\\
& \times
\bigg[
( 1 - 2 \Pmw ) \cos 2 \pi \edc^j
%
%\notag \\
%&
+ 2 \Pmw \cos  2 \pi (\edc^j + \eac) \bigg] \, .
\end{split}
\end{equation}
Although a direct addition of two different parameters $\edc^j$ and
$\eac$ has no physical meaning, we believe that, in the considered
region $\edc\simeq\eac$, it is the second term which is responsible
for apparent structure~\cite{zudovACDC07} of the differential
resistance as a function of the sum $\eac+\edc^j$, see
Fig.~\ref{fig:color}(b).

\section{Non-linear magneto-resistance beyond the bilinear response
in microwave field} \label{sec:Arbitr-Power}

In this section we extend the analysis of the magneto-transport at
$\edc \gg 1 $ to the case of arbitrary strength of  microwave radiation.
We again consider the case of relatively high microwave frequency
$\omega \gg 1/ \tautr$. The last condition ensures that the
oscillating in the variable $(t+t')/2$ part of the distribution
function is small compared to the stationary one. We therefore can
assume  that the distribution  function is homogeneous in time,
$f_{t,t'} = f_{t-t'}$.
Beyond the bilinear response, \req{K89}, obtained as the expansion
of \req{Spectrum49} in powers of $\Pmw$,
is no longer valid. We perform the time averaging of the exact collision kernel
\req{Spectrum49} over the microwave oscillation period:
\begin{equation}
\begin{split}
\label{ArbitrPower13}
\overline{\mathcal{\hat{K}}}_{t,t';\varphi}\{F(\varphi)\} = & \int\frac{d\varphi'}{2\pi}
F(\varphi')
\frac{e^{ i \pF (\vec{n}_{\varphi} - \vec{n}_{\varphi'})\vec{\zeta}^{\rm dc}_{t-t'}
}}{\tau_{\varphi-\varphi'}}
\\
&\times
\overline{e^{ i \pF (\vec{n}_{\varphi} - \vec{n}_{\varphi'})\vec{\zeta}^{\mathrm{ac}}_{t,t'}
}}\,.
\end{split}
\end{equation}
For the circular polarization the displacement due to the microwave
field $\zeta^{\rm ac}_{t,t'}$ is given by \req{K72}.
The time averaging in Eq.~\eqref{ArbitrPower13} results in
\begin{align} \label{ArbitrPower15}
\overline{e^{ i \pF (\vec{n}_{\varphi} - \vec{n}_{\varphi'})\zeta^{ac}_{t,t'} }}
= J_0\left(Q_{\varphi-\varphi'}\sin
\frac{\omega(t-t')}{ 2 }\right)\,
\end{align}
with $J_0(x)$ being the Bessel function and
\be
Q_{\varphi-\varphi'}=4 \sqrt{\Pmw
}\left|\sin\frac{\varphi-\varphi'}{2}\right|.
\ee
The time averaged collision kernel takes the form
\begin{equation}
\begin{split}
\label{ArbitrPower13b}
&  \overline{\mathcal{\hat{K}}}_{t,t';\varphi}\{F(\varphi)\}= \int\frac{d\varphi'}{2\pi}
F(\varphi')
\frac{e^{i W_{\varphi\varphi'}\cdot (t- t')} }{ \tau_{\varphi- \varphi'}}
\\
&\times
J_0\left(Q_{\varphi-\varphi'} \sin \frac{\omega(t-t')}{ 2 }\right)\,.
\end{split}
\end{equation}

The substitution of Eq.~\eqref{ArbitrPower13b} to \req{Collision15} leads to the
following expressions for the ``out'' collision term
\begin{equation}
\begin{split}
\label{ArbitrPower21}
\overline{\mathrm{St}}_{\mathrm{out}}\{ f\}_{t,t'}& =
- \int\frac{d\varphi'}{2\pi}
\frac{ f_{t,t'} }{ \tau_{\varphi - \varphi'} } %
\\
- & \int\frac{d\varphi'}{2\pi} \frac{\lambda }{ \tau_{\varphi -
\varphi'} }
J_0\left(Q_{\varphi-\varphi'} \sin \pi\eac \right) \\
\times & \left[ e^{i W_{\varphi\varphi'} \Tc }
f_{t- \Tc,t'} +
e^{ - i W_{\varphi\varphi'} \Tc }f_{t,t'- \Tc}
\right] \, .
\end{split}
\end{equation}
The ``scattering-in'' term \req{Collision28} reads
\begin{equation}
\begin{split}
\label{ArbitrPower28}
\overline{\mathrm{St}}_{\mathrm{in}} \{ f \}_{t,t'} &=
\int\frac{d\varphi'}{2\pi}
\frac{ e^{i W_{\varphi\varphi'}(t- t')} }{ \tau_{\varphi- \varphi'}}
\\
&\times
J_0\left(Q_{\varphi-\varphi'} \sin \frac{\omega(t-t')}{ 2 }\right) \\
& \times \left(f_{t,t'}-\lambda f_{t - \Tc ,t'}-\lambda f_{t
,t'-\Tc}\right) \, .
\end{split}
\end{equation}

We rewrite the kinetic equation \req{Collision32} for homogeneous in
time distribution function in the energy representation \be
\label{kineqenergy=} \wc \frac{ \partial }{ \partial \varphi }
f_{\vare;\varphi} = \mathrm {St}_{\mathrm{dis}}\{ f_{\vare;\varphi}
\}_{\vare;\varphi}\, . \ee
The full collision integral for electron scattering off disorder,
$\mathrm {St}_{\mathrm{dis}}\{ f_{\vare;\varphi}
\}_{\vare;\varphi}$, is a sum of the scattering ``in'' and ``out''
terms, given by \reqs{ArbitrPower21} and \rref{ArbitrPower28}, see
\req{KinEq23}.

We discuss the non-linear transport regime
$\edc \gg 1 $. In this case it is sufficient to consider
the ``displacement'' contribution to the non-linear current.
To calculate the ``displacement'' contribution to the electric
current, we substitute the equilibrium distribution
function~\req{ft=}
to the collision integral in \req{kineqenergy=}. Combining \reqs{ArbitrPower21} and
\rref{ArbitrPower28} and writing the result in energy
representation, we obtain
\begin{equation}
 \label{ArbitrPower31}
\mathrm{St}_\mathrm{dis}\{f_{T}(\vare)\}_{\vare;\varphi} =
\int\frac{d\varphi'}{2\pi}\frac{K_0-2\lambda \mathrm{Re} \{ e^{i
(\vare+W_{\varphi\varphi'})\Tc } K_1\}} { \tau_{\varphi-\varphi'}}.
\end{equation}
The first kernel,
\begin{equation}
K_0=J_0 \left(Q_{\varphi-\varphi'}
\sin \frac{i \omega \partial_{\vare} }{ 2 }\right)
f_{T }(\vare + W_{\varphi\varphi'})
-f_{T}(\vare),
\label{K0=}
\end{equation}
is the part of the collision integral to the zero order in
$\lambda$. In particular, it describes  the classical
Drude conductivity. Although the microwave radiation complicates
the form of $K_0$, we notice that this kernel can be represented
in terms of the series expansion in powers of
$\partial/\partial\vare$, applied to the Fermi distribution
function. The observables, such as electric current, are
determined by energy integrals. Because the integrals of all derivatives of
$f_T(\vare)$ of the second order and higher vanish, we
do not expect any effect of microwave radiation
within our model on the conductivity to the zeroth order in
$\lambda$.

The second term in \req{kineqenergy=} is
\begin{equation}
\begin{split}
\label{K1=}
K_1 = &
 J_0 \left(Q_{\varphi-\varphi'}\sin
\left[ \frac{i \omega \partial_{\epsilon} }{ 2 } -  \pi \eac \right] \right)
f_{T}(\vare + W_{\varphi\varphi'} )
\\ &
-J_0 \left(Q_{\varphi-\varphi'} \sin \pi\eac \right)
f_{T}(\epsilon ).
\end{split}
\end{equation}
In the high temperature limit, $T\gg |e|E\Rc$ and $T\gg\omega$,
\req{K1=} can be reduced to
\begin{align} \label{ArbitrPower35}
K_1 = &
\frac{\partial f_T(\vare)}{\partial\epsilon}
\Big[ J_0\left(Q_{\varphi-\varphi'} \sin \pi \eac \right)
 W_{\varphi\varphi'}
\notag \\
& + \frac{i\omega Q_{\varphi-\varphi'}}{2} J_1\left(Q_{\varphi-\varphi'} \sin \pi\eac \right)
\cos \pi\eac %
 \Big]\, .
\end{align}
Similarly to the case of the small power, we look for the correction
to the distribution function in the form
\begin{align} \label{ArbitrPower38}
\delta f (\vare,\varphi) = \delta f_{\rm cl}(\vare,\varphi)+ \lambda
\partial_{\vare}f_T
A_1 \cos \frac{ 2 \pi \vare}{ \omega_c } \cos \varphi \, .
\end{align}
Substituting $\delta f(\vare,\varphi)$ to \req{kineqenergy=}, we
find that   $\delta f_{\rm cl}(\vare,\varphi)$ is given by
a sum of \req{DF14} and higher order derivatives of $f_T(\vare)$, as discussed below \req{K0=}.
The term of the first order in $\lambda$ determines the value of
coefficient $A_1$:
\begin{equation}
\begin{split}
\label{ArbitrPower42}
A_1 & =  \frac{1}{\wc}\int\frac{d\varphi d \varphi'}{\pi^2}
\frac{\sin \varphi}{\tau_{\varphi-\varphi'}}
\\
& \times
\Bigg[
\cos\frac{2\pi W_{\varphi\varphi'}}{\wc}
J_0\left(Q_{\varphi-\varphi'} \sin \pi \eac \right)
W_{\varphi\varphi'}
 \\
&  -
 \sin\frac{2\pi W_{\varphi\varphi'}}{\wc}
 \frac{\omega Q_{\varphi-\varphi'}}{2} J_1\left(Q_{\varphi-\varphi'} \sin
 \pi\eac \right)
\cos \pi\eac\Bigg]
 \, .
\end{split}
\end{equation}

\begin{figure}[h]
\includegraphics[width=0.9\columnwidth]{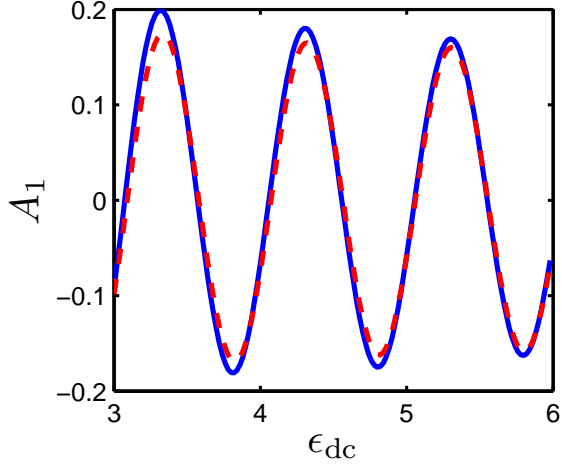}
\caption{(Color online) The function $A_1$ evaluated by (a)
numerical integration in Eq.~\eqref{ArbitrPower42}, solid line
(blue), and
(b) by the stationary phase approximation applied to
Eq.~\eqref{ArbitrPower42}, dashed line (red) for $\Pmw=8.25$,
$\eac=2.25$.
 \label{fig:stationary}}
\end{figure}

Our analysis in this section is applicable for $\edc \gtrsim
1$, when we can utilize the saddle point approximation, to perform angular
integrations in Eq.~\eqref{ArbitrPower42}. Angles $\varphi$ and
$\varphi'$ which make the phase of the integrand stationary are
$\varphi= \pm \pi/2$, $\varphi'= \mp \pi/2$ and correspond to $2R_c$
jumps along the electric field direction, see
Sec.~\ref{sec:Qualitative}. For $\sqrt{\Pmw} \lesssim \edc$ the
angular integrations in Eq.~\eqref{ArbitrPower42} results
in
\begin{align} \label{ArbitrPower50}
 & A_1 =
-\frac{ 4 }{ \pi^2 \tau_{\pi} }
\sin 2 \pi \edc
J_0\left(4 \sqrt{\Pmw } \sin \pi \eac \right)
\notag \\
&
- \frac{8 }{\pi^2 \tau_{\pi} } \frac{ \eac }{ \edc } \cos 2 \pi \edc
\cos \pi \eac
\sqrt{\Pmw } J_1\left(4 \sqrt{\Pmw } \sin \pi \eac \right)\, .
\end{align}
The result of numerical evaluation of the coefficient $A_1$, \req{ArbitrPower42}, and
its approximation, \req{ArbitrPower50}, are shown
in Fig.~\ref{fig:stationary}.
The stationary phase approximation progressively
improves as the parameter $\edc$ increases.

The non-linear contribution to the electric current is given by
\req{MagnetoR9} in term of coefficients $A_1$, \req{ArbitrPower42},
and $A_2=0$. Representing the non-linear contribution to the current
in the form of \req{deltaj=}, we write
\be\label{ArbitrPower52} F(\edc,\eac)= - \frac{\tautr}{\edc}A_1 \,
. \ee
The function $F(\edc,\eac)$ found from Eqs.~\eqref{ArbitrPower50}
and \eqref{ArbitrPower52} reads
\begin{widetext}
\begin{align} \label{ArbitrPower53}
F(\edc,\eac) = &
\frac{ 4 \tautr }{ \pi^2 \tau_{\pi} \edc } \left[
\sin 2 \pi \edc
J_0\left(4 \sqrt{\Pmw } \sin \pi \eac \right)
+ \frac{ 2 \eac }{ \edc } \cos 2 \pi \edc \cos \pi \eac
\sqrt{\Pmw } J_1\left(4 \sqrt{\Pmw } \sin \pi \eac \right) \right]\,
.
\end{align}
The oscillatory correction to the differential magneto-resistance is
obtained by substituting Eq.~\eqref{ArbitrPower53} to
Eq.~\eqref{MagnetoR41}:
\begin{align} \label{ArbitrPower57}
\frac{ \delta \rho(j) }{ \rho_D } = &
\frac{ (4 \lambda)^2 \tautr }{ \pi \tau_{\pi} } \left[
\cos 2 \pi \edc
J_0\left(4 \sqrt{\Pmw } \sin \pi \eac \right)
- \frac{ 2 \eac }{ \edc } \sin 2 \pi \edc \cos \pi \eac
\sqrt{\Pmw } J_1\left(4 \sqrt{\Pmw } \sin \pi \eac \right) \right]\,
.
\end{align}
The relation \eqref{ArbitrPower57} reduces to Eq.~\eqref{MagnetoR49}
obtained earlier in the weak power limit as expected.
\end{widetext}
\section{Discussion and Conclusions}
\label{sec:Discussion}

In this paper we have presented a comprehensive theory of an out of
equilibrium two-dimensional electron system (2DES) in a magnetic
field in the case when the disordered potential has finite
scattering amplitude on an arbitrary angle. Then, we applied this
theory to analyze electron transport in the presence of constant and
oscillating in-plane electric fields. We showed that the electric
current has an oscillating component as a function of the strength
of the constant electric field and of the frequency of an
oscillating electric field. We have investigated the position of
maxima and minima of the differential resistance on the plane of
parameters $\edc$ and $\eac$ and found a qualitative agreement with
experiments of Refs.~\ocite{zudovACDC07,zudovACDC08}.

Actual value of the differential resistance as a function of the
current through a 2DES sample depends on a number of different
scattering rates, including the full quantum scattering rate, the transport
scattering rate and the backscattering rate. Our analytical results may
be applied to the experimentally measured differential resistance to
evaluate these scattering rates and to obtain a detailed picture of
the origin and structure of the disorder in high-mobility 2DES
samples. In particular, we would like to emphasize that the behavior
of the differential resistance at large currents in experiments of
Refs.~\ocite{zudovDC06,zudovACDC07,zudovACDC08} suggests that the
disordered potential has a noticeably strong short-range component,
responsible for the finite backscattering rate.

The presence of the short range disorder also modifies the current
behavior in weak constant electric fields and at weak power of an
oscillating electric field. In particular, there are two competing
contributions to the current. One contribution originates from the
modification of the isotropic component of the electron distribution
function by electric fields,\cite{DVAMP} we refer to this
contribution as an ``inelastic'' contribution. The other
contribution is due to the modification of electron scattering rate
off disorder by electric fields, known as the displacement
contribution.\cite{ryzhii,durst03,VA03,xs03} While in smooth
disorder in weak fields the ``inelastic'' contribution dominates
in the whole range of temperatures at which the non-linear current
is expected to survive, the short range disorder may make these two
contributions comparable. As a result, the strong temperature
dependence of the non-linear current in weak fields is expected only
in samples with sufficiently weak short-range disorder.

We also studied the dependence of non-linear current on the applied
power of oscillating electric field.
The expression \eqref{ArbitrPower57} obtained in the non-linear
transport regime $\edc \gg 1$ is not limited to the small microwave
radiation power. This expression shows that in strong
microwave fields the non-linear current has a different dependence
on parameters $\eac$ and $\edc$, \req{eac_edc}, which
qualitatively differs from the corresponding expressions in the weak
power limit. We believe that the study of non-linear current at strong radiation power may
bring additional opportunities for study microscopic characteristics
of high mobility electron systems.

\begin{acknowledgments}
We thank
I. Aleiner, I. Dmitriev, A. Kamenev, B. I. Shklovskii and M. A.
Zudov for useful
  discussions.
M.K. is supported by DOE Grant No. DE-FG02-08ER46482 and BNL LDRD
Grant No. 08-002 under Contract No. DE-AC02-98CH10886 with the U. S.
department of Energy. M.G.V. is grateful to the Aspen Center for
Physics, where a part of this work was done.
\end{acknowledgments}

\appendix
\section{Calculations of the electron distribution function}
\label{App:A}

In this appendix we give the details of the calculation of the
coefficients $I$, $A_1$ and $A_2$ of  the distribution
function in the form of \req{DF15}. Substituting Eq.~\eqref{DF15} to the kinetic equation
\eqref{Collision32} with the collision terms specified by
Eqs.~\eqref{K123} and \eqref{K127} we obtain the following system of
equations
\begin{widetext}
\begin{align} \label{DF17}
I \left[ \frac{ 1 }{ \tau_{\rm ee}(\vare) } + \frac{ 1 }{ \tau_{0} }
\right] \sin \frac{ 2 \pi \vare}{ \wc } = &
\Big\langle K_{\varphi \varphi'}(\vare)
-2 \Pmw \left[
\sin^2\pi \eac \bar{K}_{\varphi \varphi'}(\vare)
+  \omega  \sin 2 \pi \eac \bar{M}_{\varphi \varphi'}(\vare)\right]
\notag \\
& + I \left[ M_{\varphi \varphi'}(\vare)
- 2 \Pmw
\sin^2\pi \eac \bar{M}_{\varphi \varphi'}(\vare) \right]
\Big\rangle
 \, ,
\end{align}
\begin{align} \label{DF23}
- \frac{ \wc A_1 }{2} \cos \frac{2 \pi \vare }{ \wc } =&
\Big\langle \sin \varphi
\big\{ K_{\varphi \varphi'}(\vare)
%\notag \\
%
% &
-  2 \Pmw\left[ \sin^2\pi \eac \bar{K}_{\varphi \varphi'}(\vare) -
\omega \sin 2 \pi \eac \bar{M}_{\varphi \varphi'}(\vare) \right] \notag \\
& + I \left[ M_{\varphi \varphi'}(\vare) - 2 \Pmw \sin^2\pi \eac
\bar{M}_{\varphi \varphi'}(\vare) \right] \big\} \Big\rangle \, ,
\end{align}
and
\begin{align} \label{DF27}
\frac{ \wc A_2 }{ 4 } = &
I \Bigg\langle \sin \varphi \cos \frac{2 \pi (\vare + W_{\varphi
\varphi'}) }{ \wc }
\left( M_{\varphi \varphi'}(\vare) - \frac{ \sin \vare }{
\tau_{\varphi - \varphi'} } \right)
%\notag \\
%
-
%&
%
\frac{ \Pmw }{ 2 } \sum_{ \pm }
\left[ \cos \frac{2 \pi (\vare + W_{\varphi \varphi'}) }{ \wc
}\left( \bar{M}_{\varphi \varphi'}(\vare)
- \frac{ \sin \vare }{ \bar{\tau}_{\varphi - \varphi'} } \right)
\right. \notag \\
& -\left.
\cos \frac{2 \pi (\vare + W_{\varphi \varphi'} \pm \omega ) }{ \wc }
\left( \bar{M}_{\varphi \varphi'}(\vare \pm \omega) - \frac{ \sin
\vare }{ \bar{\tau}_{\varphi- \varphi'} } \right) \right]
\Bigg\rangle \, .
\end{align}
\end{widetext}
where $\langle \ldots \rangle$ stands for the averaging over angular
variables $\varphi$ and $\varphi'$. Here we have introduced integral
kernels
\begin{align} \label{DF19}
K_{\varphi \varphi'}(\vare) = & - 2 \cos \frac{2 \pi (\vare +
W_{\varphi \varphi'} ) }{ \wc }
\frac{ W_{\varphi \varphi'} }{ \tau_{\varphi -\varphi'} } \notag \\
M_{\varphi \varphi'}(\vare) = & \frac{ \sin \left[ 2 \pi (\vare +
W_{\varphi \varphi'} ) / \wc \right]}{ \tau_{\varphi - \varphi'} }
\, .
\end{align}
The kernels $\bar{K}_{\varphi \varphi'}(\vare)$ and
$\bar{M}_{\varphi \varphi'}(\vare)$ are obtained from
Eq.~\eqref{DF19} by replacing $\tau \rightarrow \bar{\tau}$ as given
by \req{qualit37}.
Keeping only the energy independent part in Eq.~\eqref{DF27}
surviving the subsequent energy integration we rewrite it as
\begin{align} \label{DF28}
\frac{ \wc A_2 }{2} =
I \left\langle \sin \varphi
\left[ M_{\varphi \varphi'}(0) - 2 \Pmw \sin^2\pi \eac
\bar{M}_{\varphi \varphi'}(0) \right] \right\rangle \, .
\end{align}
For the angular averages appearing in Eqs.~\eqref{DF17},
\eqref{DF23}, and \eqref{DF28}, we have the following expressions
\begin{subequations} \label{DF29}
\begin{align}
\left\langle K_{\varphi \varphi'}(\epsilon) \right\rangle
& = \frac{ 2 e E R_c }{\pi}\gamma'(\edc)\sin \frac{ 2 \pi \epsilon
}{ \omega_c } \, ,
\end{align}
\begin{align}
\left\langle M_{\varphi \varphi'}(\epsilon) \right\rangle
& = \gamma(\edc) \sin \frac{ 2 \pi \epsilon }{ \omega_c } \, ,
\end{align}
\begin{align}
\left\langle \sin \varphi K_{\varphi \varphi'}(\epsilon)
\right\rangle
& = \frac{ e E R_c \gamma''(\edc)}{\pi^2}\cos \frac{ 2 \pi \epsilon
}{ \omega_c } \, ,
\end{align}
\begin{align}
\left\langle \sin \varphi M_{\varphi \varphi'}(\epsilon)
\right\rangle
& =  \frac{ \gamma'(\edc) }{ 2\pi } \cos \frac{ 2 \pi \epsilon }{
\omega_c } \, .
\end{align}
\end{subequations}
In Eqs.~\eqref{DF29} the function $\gamma(\edc)$ has been introduced
in Eq.~\eqref{DF30nobar}. The angular averages involving bared
functions $\bar{K}$ and $\bar{M}$ are given by Eqs.~\eqref{DF29} with
$\bar{\gamma}(\edc)$, Eq.~\eqref{DF30bar},  replacing
$\gamma(\edc)$.
Relations \eqref{DF29} allow us to perform the angular integrations in
Eqs.~\eqref{DF17}, \eqref{DF23} and \eqref{DF27} leading to the
expressions \eqref{DF31}, \eqref{DF35} and \eqref{DF41} of the main
text.
%
%

%\bibliographystyle{prsty}
%\bibliographystyle{natbib}
%\bibliography{NL2DEG}

\begin{thebibliography}{36}
\bibitem[{aty()}]{atypo}
\bibinfo{note}{Factor 2 is missed in the argument of cosine function in
  Eq.~(3.3) of \oncite{VAG}.}
\end{thebibliography}

\end{document}